\documentclass[11pt,fleqn,a4paper]{article}

\usepackage{graphicx}

\usepackage{latexsym}
\usepackage{amsfonts}
\usepackage{amssymb}
\usepackage{amsmath}
\usepackage{bbm}
\usepackage{amsthm} 

\newcommand{\be}{\begin{equation}}
\newcommand{\ee}{\end{equation}}
\newcommand{\bel}[1]{\begin{equation}\label{#1}}
\newcommand{\bea}{\begin{eqnarray}}
\newcommand{\eea}{\end{eqnarray}}
\newcommand{\bef}{\begin{figure}}
\newcommand{\ba}{\begin{array}}
\newcommand{\ball}{\begin{array}{ll}}
\newcommand{\bacl}{\begin{array}{cl}}
\newcommand{\bacll}{\begin{array}{cll}}
\newcommand{\bal}{\begin{array}{l}}
\newcommand{\bac}{\begin{array}{c}}
\newcommand{\ea}{\end{array}}
\newcommand{\N}{{\mathbb{N}}}
\newcommand{\Z}{{\mathbb{Z}}}
\newcommand{\R}{{\mathbb{R}}}
\newcommand{\E}{{\mathbb{E}}}

\newcommand{\Lcal}{{\mathcal{L}}}
\newcommand{\1}{{\mathbbm{1}}}

\newtheorem{thm}{Theorem}[section]
\newtheorem{cor}[thm]{Corollary}
\newtheorem{lem}[thm]{Lemma}
\newtheorem{prop}[thm]{Proposition}
\newtheorem{definition}{Definition}[section]

\newcommand{\bt}[2][]{\begin{thm}\label{#2}{\bf #1}\it}
\newcommand{\et}{\end{thm}}
\newcommand{\bl}[2][]{\begin{lem}\label{#2}{\bf #1}\it}
\newcommand{\el}{\end{lem}}
\newcommand{\bc}[2][]{\begin{cor}\label{#2}{\bf #1}\it}
\newcommand{\ec}{\end{cor}}
\newcommand{\bp}[2][]{\begin{prop}\label{#2}{\bf #1}\it}
\newcommand{\ep}{\end{prop}}
\newcommand{\bd}[2][]{\begin{definition}\label{#2}{\bf #1}\rm}
\newcommand{\ed}{\end{definition}}

\newcommand{\wt}{\widetilde }

\newcommand{\ind}{{\bf 1}}

  \let\d=\delta \let\e=\varepsilon 
 \let\h=\eta  \let\l=\lambda  
 \let\p=\pi  \let\r=\rho  
    \let\D=\Delta
  \let\L=\Lambda

\usepackage{color}
\newcommand{\revs}[1]{#1}
\newcommand{\revf}{} 

\begin{document}

\title{Condensation in stochastic particle systems with stationary product measures}
\author{Paul Chleboun$^{\star\dag}$, Stefan Grosskinsky$^{\star}$\\[2mm]
{\small$^{\star}$Mathematics Institute, University of Warwick, Coventry CV4 7AL, UK}\\[2mm] 
{\small$^{\dag}$Institute of Advanced Study,  University of Warwick, Coventry CV4 7AL, UK}}
\date{\today}

\maketitle

\begin{abstract}
We study stochastic particle systems with stationary product measures that exhibit a condensation transition due to particle interactions or spatial inhomogeneities. We review previous work on the stationary behaviour and put it in the context of the equivalence of ensembles, providing a general characterization of the condensation transition for homogeneous and inhomogeneous systems in the thermodynamic limit. This leads to strengthened results on weak convergence for subcritical systems, and establishes the equivalence of ensembles for spatially inhomogeneous systems under very general conditions, extending previous results which focused on attractive and finite systems. We use relative entropy techniques which provide simple proofs, making use of general versions of local limit theorems for independent random variables.
\revs{This paper is dedicated to Herbert Spohn in honour of his 65th birthday.\revf}
\end{abstract}

\section{Introduction}

Stochastic particle systems are simple models of non-equilibrium statistical mechanics, describing basic reactions and transport of particles on discrete geometries. In different applications the particles can represent various discrete or discretized degrees of freedom. In contrast to models of equilibrium statistical mechanics, they are defined by dynamical rules and in general exhibit a family of stationary measures which cannot be characterized by an energy function. We focus on stochastic lattice gas models of pure transport where the number of particles is conserved. Several basic examples have been introduced to the mathematical literature in \cite{spitzer70}, including models with and without exclusion interactions. Classical questions that have been studied include characterization of stationary measures, phase transitions and equivalence of ensembles, large scale dynamical properties and hydrodynamic limits which are summarized in several monographs including \cite{liggett,spohnbook,kipnislandim,komorowski}.

We focus on systems 
with discrete, unbounded local state space $\N =\{ 0,1,\ldots\}$, i.e. without restriction on the number of particles per site. Such models include 
zero-range processes \cite{spitzer70,liggett73,andjel82} \revs{and\revf} misanthrope processes \cite{misanthrope}, which \revs{are\revf} a large class of systems including the recently studied inclusion process \cite{giardinaetal09,giardinaetal10} \revs{and\revf} a generalized version \cite{waclawetal11}. \revs{There are various results on such models with open boundaries, but this paper is entirely focused on closed systems. We do not consider creation or annihilation of particles due to boundary reservoirs, and on finite lattices the number of particles is a conserved quantity.\revf} 
For certain geometries and particle interactions these models can exhibit a condensation phenomenon. In this case, when the particle density exceeds a critical value the system phase separates into a condensed and a homogeneous or fluid phase. The fluid phase is distributed according to the maximal invariant measure with critical density and the excess mass concentrates on a subextensive part of the lattice, constituting the condensed phase. 

Condensation can result from spatial inhomogeneities or particle interactions in spatially homogeneous systems and so far has mostly been studied for systems with stationary product measures. The first regime is addressed in \cite{evans96,krugetal96,landim96,benjaminietal96,andjeletal00} in the context of zero-range and exclusion models with disorder, \cite{ferrarisisko} covers a more general class of systems and a comprehensive review of related results on disordered systems. All rigorous results within the above references are based on coupling techniques and require attractivity of the process, whereas \cite{getal11} covers general inhomogeneities without the use of attractivity restricted to finite lattices. In these systems the condensed phase is localized on specific sites determined by geometric effects, such as slow exit rates or large incoming rates for particles. Building on first results in \cite{drouffe98,evans00}, condensation in homogeneous systems has attracted major research interest over the last years in the context of zero-range processes and related models.  In contrast to inhomogeneous systems the condensed phase is delocalized, i.e. its location is uniformly distributed on the lattice due to symmetry and therefore not accessible in the thermodynamic limit under the usual local notions of convergence. Studying the maximum as a global observable, it has been established rigorously in a series of papers \cite{jeonetal00,stefan,ferrarietal07,armendarizetal09,agl} that the condensed phase in fact concentrates on a single lattice site, covering a relatively large class of systems with stationary product measures.

The combination of inhomogeneities and interaction driven condensation has been studied in \cite{angeletal04} for a system with a single defect site and more genererally in \cite{luis1,luis2,godrecheetal12}, the latter also providing a very good account of the literature related to both cases. 
Results on homogeneous mass transport models with continuous state space can be found in  \cite{masstransport,hanney06,giardinaetal09} and references therein, and on related systems with pair-factorised stationary measures that give rise to a spatially extended condensates in \cite{evansetal06} and \cite{waclawetal09} (see also references therein). There are many further studies of applications or variations of zero-range processes and related models which we do not address here, see \cite{evansetal05,godreche07,stefan2,godrecheetal12} for complementary reviews of the literature. 

The purpose of this paper is two-fold. Firstly, we review rigorous results on condensation in \revs{closed\revf} stochastic particle systems with stationary product measures from a thermodynamic point of view, formulating them in the context of the classical approach of the equivalence of ensembles \cite{georgii}. A general information theoretic approach \cite{csiszarkoerner} provides simple proofs in terms of convergence in relative entropy. Secondly, we use this approach to derive new results on weak convergence with respect to (unbounded) local test functions for homogeneous systems. These are important to capture the nature of the condensation transition, extending classical results on bounded functions or functions with exponential moments \cite{csiszar75} which are only sufficient for models with bounded local state space such as spin systems. We further establish the equivalence of ensembles for spatially inhomogeneous models in the thermodynamic limit extending recent results in \cite{getal11} for finite lattices, and provide a detailed discussion of possible localization and delocalization of the condensed phase. These results hold under very general assumptions, in particular without requiring attractivity, and can be proved with minimal effort at the expense of providing weaker conclusions than the work cited above.

The paper is organized as follows. In Section \ref{sec:models} we give a general result (Theorem \ref{spm}) on sufficient conditions for stationary product measures, which basically summerizes various previous work in that direction \cite{misanthrope,target,masstransport,godreche07} in a relatively coherent framework. We focus on particle jump rates between sites $x$ and $y$ depending on the occupation numbers $\eta_x ,\eta_y \in\N$ on departure and target site in a factorized form. 
We close this section with a general characterization of condensation for homogeneous or inhomogeneous systems in the thermodynamic limit, and provide a connection to conepts of phase transitions in classical statistical mechanics. Section \ref{sec:hom} is devoted to spatially homogeneous systems, including a detailed discussion of static and dynamic properties of basic examples.
\revs{This section contains new results on a strengthened form of weak convergence, in particular Theorem \ref{equi} for subcritical systems.\revf} 
In Section \ref{sec:inhom} we address systems with general spatial inhomogeneities. \revs{After reviewing previous work on finite systems and discussing various examples, we present our most significant new results in Theorems \ref{subinhom} to \ref{general} on the equivalence of ensembles in the thermodynamic limit.\revf}
Attempting to give a fairly complete picture of rigorous results on condensation in \revs{closed\revf} stochastic particle systems, we give a short account of further work in the discussion in Section \ref{sed:discussion}, including refined scaling limits at the critical density, systems with size-dependent parameters and the dynamics of condensation.

\section{Models and notation\label{sec:models}}

\revs{In this section we introduce notation and summarize previous results on stationary product measures, the equivalence of ensembles and connections to condensation.\revf}

\subsection{Definition of the dynamics\label{sec:defdyn}}

We consider a family of lattice gases, which are continuous-time Markov process with state space $X_\Lambda =\N^\Lambda$, where $\Lambda$, called the lattice, can be any finite \revs{or countably infinite\revf} set. Configurations are denoted by $\eta =(\eta_x :x\in\Lambda )$, where $\eta_x \in\N$ is the number of particles at site $x\in\Lambda$. 
\revs{We focus on closed system in which particles are not created or annihilated and their number is locally conserved.\revf}
The dynamics are given by the generator
\bea\label{gene}
\Lcal f(\eta )=\sum_{x,y\in\Lambda} p(x,y)\, u_x (\eta_x )\, v_y (\eta_y )\big( f(\eta^{xy})-f(\eta )\big)\ ,
\eea
with the usual notation $\eta^{xy}_z :=\eta_z-\delta_{z,x} +\delta_{z,y}$ for a configuration where one particle has moved from site $x$ to $y$. The purely spatial part of the jump rates, $p(x,y)\geq 0$ are transition rates of a single random walker on $\Lambda$ with $p(x,x)=0$, which we assume to be irreducible to avoid hidden conservation laws. The interaction part of the jump rates is given by functions $u_x ,v_x :\N\to [0,\infty )$ for each $x\in\Lambda$, which should satisfy
\bea
u_x (n)&=&0\quad\mbox{if and only if}\quad n=0\ ,\nonumber\\
v_x (n)&>&0\mbox{ for all }n\geq 0\quad\mbox{and}\quad v_x (0)=1\ ,
\eea
for all $x\in\Lambda$. Positivity ensures that there are no degeneracies or absorbing states, and the normalization of $v_x (0)$ is just a convenient choice and no restriction, since it can be absorbed in rescaling the rates \revs{$u_x (n)$.\revf}

With these assumptions the number of particles is the only conserved quantitiy and for finite lattices of size $|\Lambda |=L$ the process is irreducible on the subsets
\bea
X_{\Lambda ,N} =\big\{\eta\in X :\Sigma_\Lambda (\eta )=N\big\}\quad\mbox{for each}\quad N\in\N\ ,
\eea
where we use the shorthand $\Sigma_\Lambda (\eta )=\sum_{x\in\Lambda} \eta_x$. On $X_{\Lambda ,N}$ the process is a finite state, irreducible Markov chain, and is therefore ergodic with a unique stationary measure $\pi_{\Lambda ,N}$. 
In this case the generator is defined for all continuous functions $f\in C(X_{\Lambda ,N})$. Examples of processes with such dynamics include
\begin{itemize}
	\item zero-range processes (ZRP) \cite{spitzer70}: $u_x$ arbitrary\ ,\quad $v_x \equiv 1$\ ;
	\item target process (TP) \cite{target}: $u_x (n)=1-\delta_{n,0}\ ,\ \  v_x (n)= v(n)$ arbitrary\ ;
	\item inclusion processes (IP) \cite{giardinaetal09,giardinaetal10}: $u_x (n)=n\, ,\ v_x (n)=d+n\, ,\ d>0$\ ;
	\item explosive condensation model (ECP) \cite{waclawetal11}:\\
	$u_x (n)=v(n)-v(0)\ ,\quad v_x (n)=v(n)=(d+n)^\gamma\ ,\quad d,\gamma >0$\ .
\end{itemize}
In the ZRP particles exhibit only an on-site, zero-range interaction, while in the IP and ECP particles can be attracted by a large occupation number on the target site, which is also true for the TP for increasing $v(n)$. In addition to this interaction, particles perform independent random walks with rate $d\, p(x,y)$ in the IP, whereas in the ECP jump rates depend superlinearly on the occupation on departure sites leading to a repulsive interaction if $\gamma >1$, which is the interesting case for this model. For $\gamma =1$ the ECP is equivalent to the IP. Note that the rates in their original form as given above do not obey the normalization $v_x (0)=1$ for TP, IP and ECP, but this can easily be achieved by rescaling $u_x$ by $v_x(0)$, $d$ and $d^\gamma$, respectively.

The family of processes (\ref{gene}) has some overlap with misanthrope processes \cite{misanthrope}, which were originally defined with translation invariant lattices and jump rates $p(x,y) =q(y-x)$, and a more general interaction part of the jump rates given by a function $g(\eta_x ,\eta_y )$. The processes are known to exhibit stationary product measures if the rates fulfill
\bea
\frac{g(n,m)}{g(m+1,n-1)} =\frac{g(n,0)\, g(1,m)}{g(m+1,0)\, g(1,n-1)}\quad\mbox{for all}\quad n\geq 1,\, m\geq 0\ ,\nonumber
\eea
and, in addition, either $q(z)=q(-z)$ for all $z\in\Lambda$ or $g(n,m)-g(m,n)=g(n,0)-g(m,0)$ for all $n,m\geq 0$.
All translation invariant examples we will consider are in fact special misanthrope models, but we are explicitly also interested in spatially inhomogeneous cases. Models of type (\ref{gene}) are attractive, i.e. they preserve stochastic order in time, if and only if $u_x$ are increasing and $v_x$ are decreasing functions. This is analogous to well known results for misanthrope models \cite{misanthrope,saada}. Here we are particularly interested in condensation phenomena, which for homogeneous systems so far have only been observed if this condition is violated and the model is not attractive. It is an interesting open question whether non-attractiveness is in fact a necessary condition for condensation in homogeneous systems. 

To construct the dynamics on infinite lattices, such as $\Lambda =\Z$, further assumptions are necessary \cite{liggett73}. Since the local state space $\N$ and therefore also $X$ is non-compact, the usual construction using Feller semigroups and continuous cylinder functions $f$ to define the generator (see \cite{liggett}, Chapter I) does not apply. In \cite{andjel82} a construction is given for a spatially homogeneous ZRP ($u_x \equiv u$) using Lipschitz functions $f$ on a restricted state space to limit the growth of $\eta_x$ as $|x|\to\infty$, under the additional assumptions that $p$ is of finite range and the jump rates $g(n)$ are uniformly bounded by a linear function. This has recently been generalized in \cite{balazsetal07} to superlinear growth rates for attractive systems. In this paper we are not interested in infinite lattices directly, but take the statistical mechanics approach and study observables of large, finite systems as their size tends to infinity.

\subsection{Stationary product measures\label{sec:statmeas}}

In the following we will give sufficient conditions for processes (\ref{gene}) to exhibit stationary product measures, which we write as
\bea\label{pm}
\nu_\phi^\Lambda [d\eta ]=\prod_{x\in\Lambda} \bar\nu_\phi^x (\eta_x ) d\eta
\eea
defined by product densities w.r.t.\ the product counting measure $d\eta$ on $X_\Lambda$. The marginals turn out to have the form
\bea\label{pmarg}
\nu_\phi^x [\eta_x =n]=\bar\nu_\phi^x (n)=\frac{1}{z_x (\phi )}\, w_x (n)\, (\lambda_x \phi )^n 
\eea
with normalization (or partition function)
\bea\label{pfun}
z_x (\phi )=\sum_{n=0}^\infty w_x (n)\, (\lambda_x \phi )^n \ .
\eea
Here $(\lambda_x :x\in\Lambda )$ is a harmonic function
\bea\label{harm}
\sum_{x\in\Lambda} \Big(\lambda_x \, p(x,y)-\lambda_y p(y,x)\Big)=0\quad\mbox{for all }y\in\Lambda\ ,
\eea
corresponding to a (not necessarily normalized) stationary distribution of a single random walker with transition rates $p(x,y)$. Since we assume $p(x,y)$ to be irreducible, on finite lattices $\Lambda$ this is in fact unique up to normalization and strictly positive. The weights $w_x$ are given by
\bea\label{weight}
w_x (n)=\prod_{k=1}^n \frac{v_x (k-1)}{u_x (k)}\ ,\quad x\in\Lambda \ ,
\eea
encoding the interaction of the particles provided through the functional forms of $u_x$ and $v_x$.

Since the number of particles is a conserved quantity, the measures are indexed by a fugacity parameter $\phi\geq 0$ controling the average number of particles per site
\bea\label{rphi}
R_x (\phi )=\nu_\phi^x (\eta_x )=\frac{1}{z_x (\phi )} \sum_{n=0}^\infty n\, w_x (n)\, (\lambda_x \phi )^n\ ,
\eea
which is a strictly increasing function with $R_x (0)=0$. Here and in the following we use the standard notation $\mu (f)$ to denote the expectation of a function $f$ under a measure $\mu$. Since the normalization $z_x (\phi )$ 
is a generating function, the density can also be computed as $R_x (\phi )=\phi\,\partial_\phi \log z_x (\phi )$. Existence of the product measure (\ref{pm}) obviously requires $z_x (\phi )<\infty$ for all $x\in\Lambda$, and we denote by
$$
D_\phi^\Lambda =\big\{\phi\geq 0\, :\, z_x (\phi )<\infty\mbox{ for all }x\in\Lambda\big\}
$$
the domain of definition. Since $z_x (\phi )$ is a power series in $\phi$, the domain of each marginal $\nu^x_\phi$ is actually of the form $D^\Lambda_\phi =[0,\phi_c^x )$ or $[0,\phi_c^x ]$ where
$$
\phi_c^x =\big(\lambda_x \limsup_{n\to\infty} w_x (n)^{1/n} \big)^{-1}
$$
is the radius of convergence of $z_x (\phi )$. The domain of the product measure is then
\bea\label{domain}
D^\Lambda_\phi =[0,\phi_c^\Lambda )\quad\mbox{or}\quad [0,\phi_c^\Lambda ]\quad\mbox{where}\quad \phi_c^\Lambda =\inf_{x\in\Lambda} \phi_c^x \ .
\eea
Whether or not the right boundary is part of the domain depends on the particular example and is related to the condensation phenomenon which is discussed later in detail. For non-empty $D_\phi^\Lambda$ we need $\phi_c^\Lambda >0$. A sufficient condition is for example that for all $x\in\Lambda$
\bea\label{gencond}
\frac{1}{n}\log w_x (n)=\frac{1}{n}\sum_{k=1}^n \log\frac{v_x (k-1)}{u_x (k)}\to\alpha_x \in [-\infty ,\infty )
\eea
as $n\to\infty$, and
\bea\label{lambound}
\lambda_x ,\ \alpha_x \leq C\quad\mbox{are uniformly bounded above for all }x\ ,
\eea
leading to $\phi_c^x =e^{-\alpha_x} /\lambda_x \geq e^{-C}/C$. Uniform boundedness is clear on fixed lattices but a non-trivial condition in the thermodynamic limit, and (\ref{gencond}) obiously holds whenever $v_x (n-1)/u_x (n)$ has a finite limit for all $x$ as $n\to\infty$.

\bt[Stationary product measures\\]{spm}
The processes with generator (\ref{gene}) have stationary product measures $\nu_\phi^\Lambda$ of the form (\ref{pm}), provided that one of the following conditions holds:
\begin{enumerate}
	\item $v_x (n)\equiv 1$ for all $x\in\Lambda$, $n\geq 0$ (zero-range dynamics).
	\item The harmonic function $\lambda_x$ (\ref{harm}) fulfilles the detailed balance relation
	\bea\label{detbal}
	\lambda_x p(x,y) =\lambda_y p(y,x)\quad\mbox{for all }x,y\in\Lambda\ .
	\eea
	In this case the measure is in fact reversible for the dynamics (\ref{gene}).
	\item Incoming and outgoing rates $p$ are the same for each site, i.e.
	\bea\label{dstoch}
	\sum_{y\in\Lambda} p(x,y)=\sum_{y\in\Lambda} p(y,x)\quad\mbox{for all }x\in\Lambda\ ,
	\eea
	and $v_x =v,\, u_x =u$ are independent of $x$ and fulfill
	\bea\label{homcon}
	u(n)\, v(m)-u(m)\, v(n)=u(n)-u(m)\quad\mbox{for all }n,m\geq 0\ .
	\eea
	In this case the measure is homogeneous with $x$-independent marginals.
\end{enumerate}
\et

Cases 1 and 3 are known from the literature of zero-range models \cite{spitzer70,andjel82,kipnislandim} and misanthrope models \cite{misanthrope} in a slight reformulation. Case 2 is a straightforward extension to a proof for inclusion processes given in \cite{getal11} \revs{including one of the authors\revf}, which is based on a classical result on the exclusion process \cite[Theorem VIII.2.1]{liggett}. Cases 2 and 3 have also been discussed in \cite{target} in the context of the target process. Another recent model covered by case 3 is the explosive condensation model studied in \cite{waclawetal11} for totally asymmetric dynamics on a one-dimensional torus, and the theorem also holds for other geometries. For completeness we give a short summary of the main steps in the proof.

\begin{proof} We have to show for expected values w.r.t. $\nu_\phi$ that
\be\label{lzero}
\nu^\Lambda_\phi (Lf)=\sum_{\eta\in\Omega} \sum_{x,y\in\Lambda} p(x,y) u_x (\eta_x )v_y (\eta_y ) (f(\eta^{x,y})-f(\eta))\bar\nu^\Lambda_\phi (\eta )=0
\ee
for all local functions $f$. For fixed $x,y$ we get after a change of variable
$$
\sum_{\eta\in\Omega} u_x (\eta_x )v_y (\eta_y ) f(\eta^{x,y}) \bar\nu^\Lambda_\phi (\eta ) =\sum_{\eta\in\Omega} u_x (\eta_x {+}1 )v_y (\eta_y {-}1) f(\eta ) \bar\nu^\Lambda_\phi (\eta^{y,x} )\ .
$$
The form (\ref{pmarg}) and (\ref{weight}) of the marginals implies that for all $x,y\in\Lambda$ and $n\geq 0$, $k\geq 1$
\bea\label{recurs}
\bar\nu_\phi^x (n{+}1)\,\bar\nu_\phi^y (k{-}1)\, u_x (n{+}1)\, v_y (k{-}1) =\bar\nu_\phi^x (n)\,\bar\nu_\phi^y (k)\, u_y (k)\, v_x (n)\,\frac{\lambda_x}{\lambda_y}\ .
\eea
It is easy to check that boundary terms in the sums vanish consistently, and we do not consider them in the following. Plugging this into (\ref{lzero}) we get for the right-hand side
\be\label{final1}
\sum_{\eta\in\Omega} f(\eta )\bar\nu^\Lambda_\phi (\eta )\sum_{x,y\in\Lambda} p(x,y)\left( u_y (\eta_y )v_x (\eta_x )\frac{\lambda_x}{\lambda_y}- u_x (\eta_x )v_y (\eta_y )\right)\ ,
\ee
and exchanging the summation variables $x\leftrightarrow y$ in the first part of the sum leads to
$$
\nu^\Lambda_\phi (Lf)=\sum_{\eta\in\Omega} f(\eta )\bar\nu_\phi (\eta )\sum_{x\in\Lambda} \frac{u_x (\eta_x )}{\lambda_x} \sum_{y\in\Lambda} v_y (\eta_y )\big( p(y,x)\lambda_y {-}p(x,y)\lambda_x\big)\ .
$$
This clearly vanishes in the first two cases and analogously to the above argument one can show that in the second case detailed balance implies
$$
\nu_\phi^\Lambda (f Lg)=\nu_\phi^\Lambda (g Lf)\quad\mbox{for local functions }f,g\ ,
$$
i.e. $\nu_\phi^\Lambda$ is reversible. \\
For the homogeneous case 3, we can use $\lambda_x \equiv 1$ and (\ref{homcon}) in (\ref{final1}) to get
$$
\nu_\phi^\Lambda (Lf)=\sum_{\eta\in\Omega} f(\eta )\bar\nu_\phi (\eta )\sum_{x\in\Lambda} u_x (\eta_x ) \sum_{y\in\Lambda} \big( p(y,x) -p(x,y)\big)\ ,
$$
which vanishes due to (\ref{dstoch}).
\end{proof}

\noindent\textbf{Remarks.}
\begin{itemize}
	\item Note that in many instances the above measures can be extended to infinite lattices in a generic way, even if existence of the dynamics of the process is not guaranteed. If the the dynamics exist the measures are stationary for the limiting dynamics, and since the harmonic functions are no longer unique, there might even be a larger family of stationary product measures for a given process.
	\item The above result also applies if $v_x (k)=0$ for all $k\geq K$, $x\in\Lambda$ for some $K\in\N$, i.e. for exclusion processes \cite{liggett} or $K$-exclusion type models with restricted state space $\{ 0,\ldots ,K\}^\Lambda$ (cf. \cite[Section II.2.4]{spohnbook} and references therein).
	\item For systems with open boundaries the theorem can be generalized directly to special cases, where each boundary can be described consistently by a single auxiliary external site. Precisely, let $\Delta$ be the set of external sites $e$, then in addition to the bulk dynamics given in (\ref{gene}) the generator has the additional terms
	\bea
	\sum_{e\in\Delta} \sum_{y\in\Lambda} p(e,y) \alpha_e v_y (\eta_y )\big( f(\eta^{+y} )-f(\eta )\big) +\nonumber\\
	\sum_{x\in\Lambda} \sum_{e\in\Delta} p(x,e) \beta_e u_x (\eta_x )\big( f(\eta^{-x} )-f(\eta )\big)
	\eea
	for creation and annihilation of particles, with the obvious notation $\eta^{\pm x}_z=\eta_z \pm\delta_{z,x}$. Consistency means, that there exists a fugacity $\phi^*\in D_\phi$ such that the total system including the auxiliary sites is a closed boundary system with a product measure $\nu_{\phi^*}^{\Lambda\cup\Delta}$ and a particular harmonic function $(\lambda_x :x\in\Lambda\cup\Delta )$. The creation and annihilation rates then have to be expectations w.r.t. marginals on external sites, i.e.
	\bea
	\alpha_e =\nu_{\phi^*}^{\Lambda\cup\Delta} (u_e )\quad\mbox{and}\quad \beta_e =\nu_{\phi^*}^{\Lambda\cup\Delta} (v_e )\ .
	\eea
	This is in general not restrictive in the reversible case (2) and for zero-range dynamics (1) since the functions $u_e$ and $v_e$ can be chosen essentially arbitrarily, but imposes restrictions in the homogeneous case (3). If we assume irreducibility of $p(x,y)$ on the extended finite lattice the scaled harmonic function $\phi\lambda_x$ is unique and there is at most one product measure for such open boundary systems.
	\item If the above consistency relations are not fulfilled the stationary measures are in general not of product form, as is well known e.g. for the one-dimensional simple exclusion process where the correlation structure can be described using a matrix product formulation following work in \cite{derridaetal93}. It is an interesting open question whether this technique also applies to other models mentioned in Section \ref{sec:defdyn} such as the IP or ECP, and if there is a connection with condensation in open boundary systems which has so far only been studied for the ZRP in \cite{levineetal05},
\end{itemize}

%



\subsection{Condensation and equivalence of ensembles\label{sec:conen}}

Although the result on stationary product measures applies in more generality, for the rest of this paper we are interested in closed finite systems and their scaling limits, where the number of particles is the only conserved quantity and there is no restriction on the number of particles per site. We will mostly be interested in stationary properties, which reduces to a study of $L=|\Lambda |$ independent random variables $\eta_x$ with distribution $\nu_\phi^\Lambda$ (\ref{pm}) and marginals (\ref{pmarg})
$$
\nu_\phi^\Lambda [\eta_x =n]=\bar\nu_\phi^x (n)=\frac{1}{z_x (\phi )}\, w_x (n)\, (\lambda_x \phi )^n \ . 
$$
In the following we will further assume that the weights $w_x (n)>0$ are sub-exponential in the sense that
\bea\label{subass}
\frac{w_x (n+1)}{w_x (n)}=\frac{v_x (n)}{u_x (n-1)}\to 1\quad\mbox{as}\quad n\to\infty\ .
\eea
This is not a restriction, since any exponential part can be absorbed in a redefinition of $\lambda_x$, the harmonic nature of which (\ref{harm}) is no longer important at this stage. The only case not covered is if the weights $w_x$ have super-exponential decay, but then $\phi_c^x =\infty$ for such sites and they do not contribute to condensation, so we do not consider this case. The most important aspect remaining from the dynamical origin of these measures is the fugacity $\phi$, the dual parameter to the conserved quantitity, which indexes the family of product measures. 

While (\ref{subass}) is sufficient on finite lattices, we need to impose some uniformity in $x$ to get sufficient control on inhomogeneous systems in the thermodynamic limit. We assume that there exist functions $w_-$ and $w_+$ such that
\bea\label{wbound}
& w_- (n)\leq w_x (n)\leq w_+ (n)&\quad \mbox{for all }n\geq 0,\ x=1,2,\ldots\quad\mbox{and}\nonumber\\
&\displaystyle\frac{w_+ (n+1)}{w_+ (n)}\, ,\ \frac{w_- (n+1)}{w_- (n)}\to 1&\quad \mbox{as}\quad n\to\infty\ ,
\eea
Together with the assumption that
\bea\label{lambound2}
\lambda_x \leq C\quad\mbox{uniformly for all }x=1,2,\ldots\ ,
\eea
this implies and replaces earlier conditions (\ref{gencond}) and (\ref{lambound}), and ensures a non-empty domain of definition as we discuss below. We note that (\ref{wbound}) is a relatively weak assumption, $w_-$ could be a decreasing and $w_+$ an increasing function with sub-exponential tails, which are otherwise arbitrary. 


The product measures also provide explicit formulas for the canonical measures $\pi_{\Lambda ,N}$ on the irreducible subsets $X_{\Lambda ,N}$. Since the number of particles $\Sigma_\Lambda$ is conserved under the dynamics, the conditional measures $\nu_\phi^\Lambda \big(d\eta\big| X_{\Lambda ,N}\big)$ are also stationary, 
and since the process is ergodic on $X_{\Lambda ,N}$ these conditional measures are equal to $\pi_{\Lambda ,N}$ and independent of the fugacity $\phi$. Choosing $\phi =1$ for simplicity we can write
\bea\label{can}
\pi_{\Lambda ,N} [d\eta ]=\nu_1^\Lambda \big[ d\eta\big| X_{\Lambda ,N}\big] =\frac{1}{Z_{\Lambda ,N}} \prod_{x\in\Lambda} w_x (\eta_x )\,\lambda_x^{\eta_x} \, d\eta
\eea
with $Z_{\Lambda ,N} =\nu_1^\Lambda [X_{\Lambda ,N}]$ as normalization.

As usual (see \cite{liggett}, Proposition I.1.8), the set of all stationary measures of the models (\ref{gene}) is a convex subset of measures on $X_\Lambda$. On finite lattices $\Lambda$ the canonical measures $\pi_{\Lambda ,N}$ are the extreme points for this set, and the (grand-canonical) product measures $\nu_\phi^\Lambda$
can be written as convex combinations
\bea
\nu_\phi^\Lambda =\sum_{N\in\N} \nu_\phi^\Lambda [X_{\Lambda ,N}]\,\pi_{\Lambda ,N} \ ,
\eea
and are therefore not extremal. On finite lattices there are no other extremal measures than the canonical ones, and the full set of stationary distributions is given by their convex hull. On infinite lattices the situation is more complicated. In spatially homogeneous systems the grand-canonical measures are extremal, but there may be more non-homogeneous extremal measures analogous to the so-called blocking measures for exclusion processes (see e.g. \cite[Chapter VIII]{liggett}).

In the thermodynamic limit
\bea\label{thermo}
L=|\Lambda |,N\to\infty\quad\mbox{such that}\quad N/L\to\rho\geq 0
\eea
the grand-canonical measures (with simple product structure) are usually expected to provide a good approximation to the sequence of canonical measures, which is called the equivalence of ensembles in statistical mechanics. One convenient way of quantifying the distance between the two distributions is relative entropy (see e.g. \cite{csiszarkoerner}, Chapter I.3). For two measures $\mu_1 ,\mu_2$ on a countable space $\Omega$ it is defined as
\bea\label{relentdef}
H(\mu_1 ;\mu_2 )=\left\{\bacl \sum_{\omega\in\Omega} \mu_1 (\omega )\log\frac{\mu_1 (\omega )}{\mu_2 (\omega )}&,\ \mathrm{if}\ \mu_1 \ll\mu_2\\ \infty &,\ \mathrm{otherwise} \ea\right.\ ,
\eea
\revs{where we use the convention $0\log 0 = 0$.\revf}
It only takes finite values if $\mu_1$ is absolutely continuous w.r.t. $\mu_2$, which means that for all measurable events $A$, $\mu_2 (A)=0$ implies that $\mu_1 (A)=0$. In this case the Radon-Nikodym derivative $h(\omega )=\frac{d\mu_1}{d\mu_2}(\omega )=\frac{\mu_1 (\omega )}{\mu_2 (\omega )}$ exists (taking a simple form in the discrete case), and the relative entropy can be written as
\bea
H(\mu_1 ;\mu_2 )=\sum_{\omega\in\Omega} \mu_2 (\omega )\, h(\omega )\log h(\omega )=\mu_2 \big( h\log h \big)\ .
\eea
Note that the relative entropy is not symmetric and therefore not a metric, but if $\mu_1$ and $\mu_2$ are probability measures it is non-negative and vanishes if and only if $\mu_1 =\mu_2$.

\begin{figure}
\includegraphics[width=0.49\textwidth]{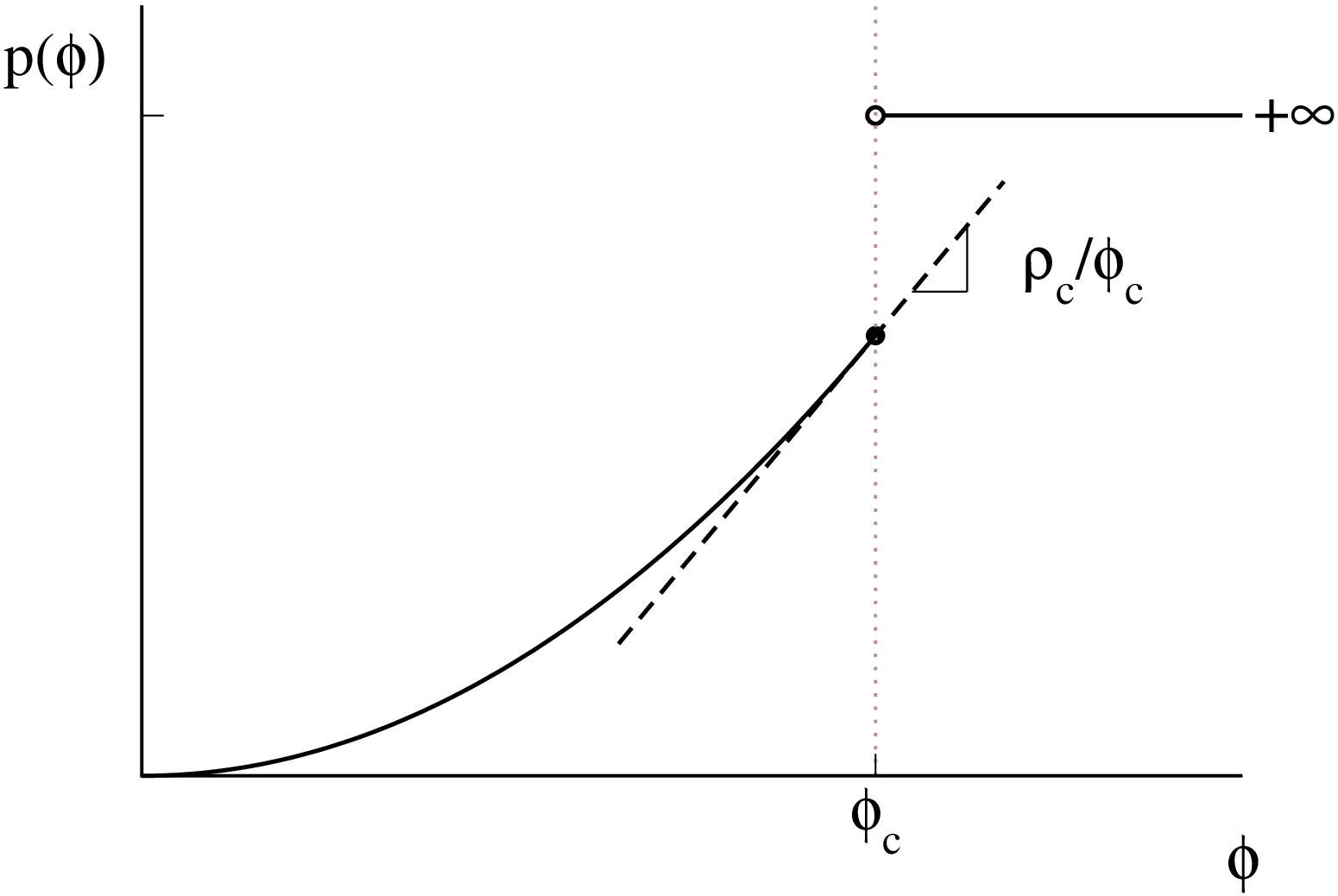}
\hspace{1ex}
\includegraphics[width=0.49\textwidth]{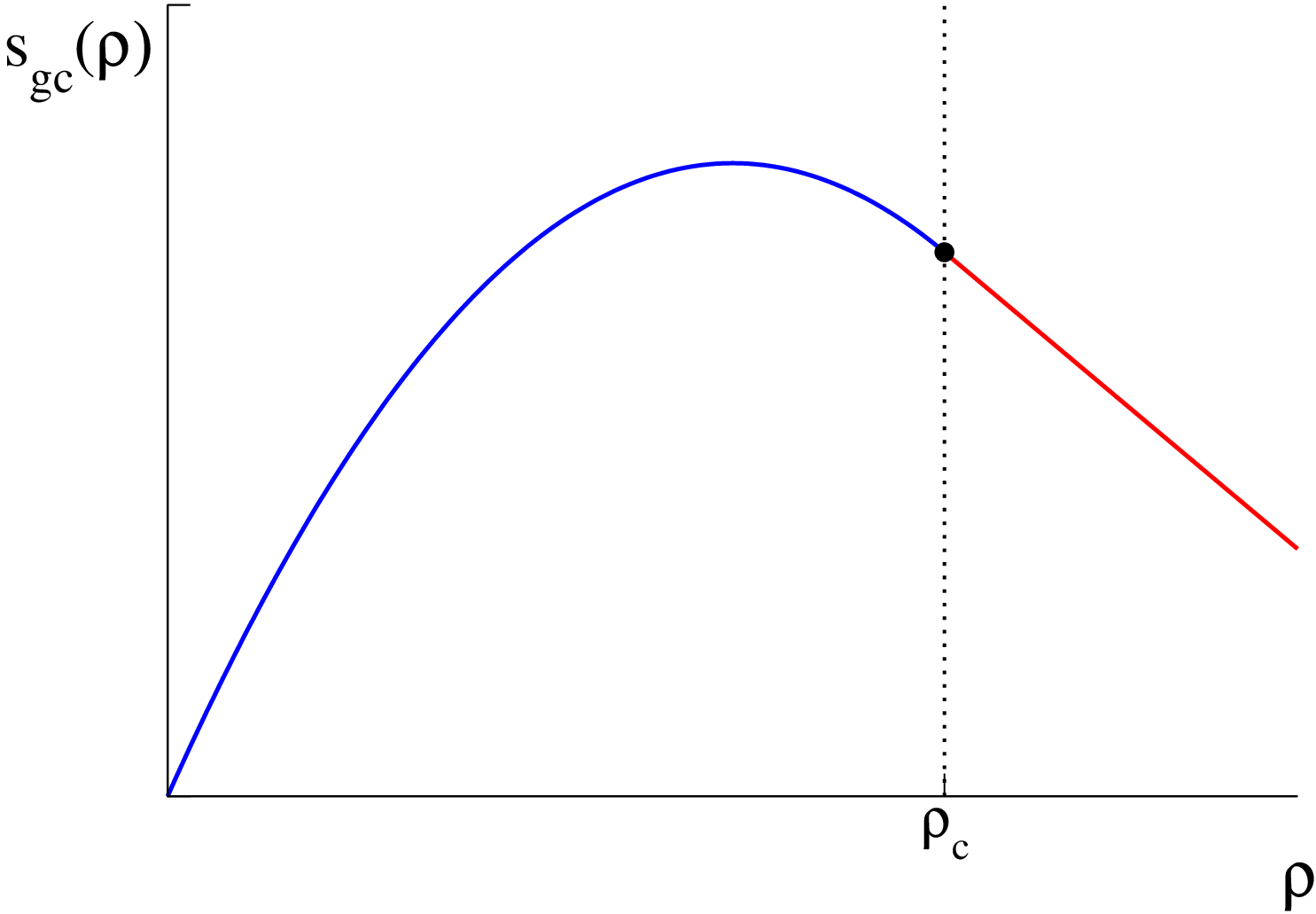}
\caption{\label{fig1}
Sketch of the pressure $p(\phi )$ (left) and its Legendre transform, the grand-canonical entropy $s_{gc} (\rho )$ (right) for a condensed system with $\rho_c <\infty$. 
\revs{$p(\phi)$ is convex and therefore Lipschitz on the interior of its domain, but not necessarily left-continuous at $\phi_c$.\revf}
By (\ref{limdens}), $\rho_c =\lim_{\phi\nearrow\phi_c} \phi \, p'(\phi )$ and simply $\rho_c =\phi_c \, p'(\phi_c )$ in case $D_\phi =[0,\phi_c ]$ as depicted in this example. In both cases the finite (limiting) slope at $\phi_c$ leads to a linear part in $s(\rho )$ for $\rho\geq\rho_c$.
}
\end{figure}

In our case, since the canonical measures are conditioned versions of the grand-canonical ones (\ref{can}), it is easy to see (cf. \cite{csiszarkoerner,stefan}) that the specific relative entropy normalized by the system size can be written as
\bea
\frac1L H(\pi_{\Lambda ,N};\nu_\phi^\Lambda )&=&-\frac1L \log\nu_\phi^\Lambda \big(\Sigma_\Lambda =N\big) \label{ldcon}\\
&=&\frac1L\log z^\Lambda (\phi )-\frac{N}{L}\log\phi -\frac1L\log Z_{\Lambda ,N} \ .\label{thermocon}
\eea
The first line provides a formulation in terms of typical or large deviations for the product measure $\nu_\phi^\Lambda$, which we will use in the following to show that the quantity vanishes in the thermodynamic limit. The second form provides a connection to thermodynamics which is not essential for our approach but we discuss it briefly for completeness, for further details and references see e.g. \cite{touchette}. In the thermodynamic limit the first term is the pressure $p(\phi )$ of the grand-canonical system (sometimes also called Gibbs free energy), and the last term is the entropy density $s(\rho )$ of the canonical system,
\bea
p(\phi )&=&\lim_{L\to\infty} \frac1L\log z^\Lambda (\phi ) =\lim_{L\to\infty} \frac1L\sum_{x\in\Lambda} \log z_x (\phi )
\ ,\label{pressure}\\
s(\rho )&=&\lim_{L\to\infty} \frac1L\log Z_{\Lambda ,N}\ .\label{entropy}
\eea
The grand-canonical entropy $s_{gc}(\rho )$ is given by the Legendre transform of the pressure, and taking the infimum over $\phi$ we get in the thermodynamic limit
\bea
\underbrace{\inf_{\phi \in D_\phi} \big( p(\phi )-\rho\log\phi\big)}_{:=s_{gc} (\rho )} -s(\rho )=\inf_{\phi \in D_\phi}\lim_{L\to\infty} \frac1L H(\pi_{\Lambda ,N};\nu_\phi^\Lambda )\ .
\eea
So if the specific relative entropy vanishes in the limit under the optimal choice of $\phi$ on the domain $D_\phi$, which we discuss below, we have equivalence of ensembles in the sense of thermodynamic functions $s_{gc}(\rho ) =s(\rho )$.
\revs{The grand-canonical entropy density $s_{gc}(\rho)$ is always concave by definition, and it turns out that it is also strictly concave for $\rho < \rho_c$.}
By general arguments, equivalence certainly holds when the grand-canonical entropy $s_{gc} (\rho )$ is \emph{strictly} \revs{concave\revf}, which can therefore be understood directly from the pressure \revs{as is shown in \cite{paulthesis,paulinprep} by one of the authors\revf}. 
\revs{The connection to particle density is explained in more detial below and illustrated in Figure \ref{fig1}.\revf}

%
%

For the family of grand-canonical measures $\nu_\phi^\Lambda$ we define the limiting particle density
\bea\label{limdens}
R (\phi ):=\lim_{L\to\infty} \frac{1}{L} \sum_{x\in\Lambda} R_x (\phi ) =\phi \, p'(\phi )\ ,
\eea
which can be written in terms of the pressure since this is a moment generating function. In the following we consider $D_\phi$ to be the domain of this function rather than the pressure, both domains are again intervals of the form $[0,\phi_c )$ or $[0,\phi_c ]$ and can disagree only at their right boundary. In fact
\bea\label{limdom}
\phi_c =\lim_{L\to\infty} \phi_c^\Lambda >0\ ,
\eea
is simply given by the limit of finite systems (\ref{domain}), which is a consequence of the uniformity assumption (\ref{wbound}) without which $\phi_c$ might be striclty smaller than the limit. $\phi_c >0$ follows from condition (\ref{lambound2}). As a limit of convex functions $p$ is convex, non-negative and $p(0)=0$, and therefore $R$ is monotone increasing with $R(0)=0$. We further assume that
\bea\label{nondeg}
p\mbox{ is strictly convex },\ R\mbox{ is strictly monotone, continuous on }D_\phi\ .
\eea
This excludes systems that become degenerate in the thermodynamic limit, for example inhomogeneous systems with $\lambda_x \to 0$ as $x\to\infty$ for which $p$ and $R$ would vanish due to (\ref{wbound}). Also condensation in the homogeneous SIP with system size dependent parameters studied in \cite{getal11} is excluded, where the pressure and the density vanish on $D_\phi$ as well. A simple condition to ensure (\ref{nondeg}) would be to assume $\lambda_x \geq C$ to be bounded from below, but this is far from necessary. In explicit examples (\ref{nondeg}) is usually easy to check and often holds, unless in special cases.


\bd{condensation}
The \textit{critical density} $\rho_c \in [0,\infty ]$ is defined as
\bea\label{rhoc}
\rho_c :=\lim_{\phi\nearrow\phi_c} R (\phi )\quad\mbox{with $R(\phi )$ as in (\ref{limdens})}\ ,
\eea
and the system exhibits \textit{condensation} if $\rho_c <\infty$.
\ed

As is illustrated in Figure \ref{fig1}, $\rho_c /\phi_c$ is the slope of the pressure $p$ \revs{as $\phi \to \phi_c$\revf} and $\rho_c <\infty$ leads to a linear part in the Legendre transform, confirming that the density is the appropriate observable to characterize the condensation transition. It is clear that $\phi_c <\infty$ is a necessary condition for condensation, see e.g. \cite[Lemma II.3.3]{kipnislandim} for a proof in a special case. For example, if the stationary weights had super-exponential decay, as is e.g. the case for independent random walkers where the $\eta_x$ are i.i.d. Poisson random variables, we have $\phi_c =\infty$ and necessarily $\rho_c =\infty$ and there is no condensation.

The above characterization of condensation works well in the thermodynamic limit for homogeneous and inhomogeneous systems, which will be explored in more detail in the next two sections. It also works for systems with size-dependent parameters, which we shortly discuss in Section \ref{sec:further}. For other scaling limits, such as $N\to\infty$ on a fixed lattice $\Lambda$, the above definition has to be adapted and we discuss previous results in this case in Proposition \ref{fininhom} for inhomogeneous and in Section \ref{sec:further} for homogeneous systems.


\section{Condensation in homogeneous systems\label{sec:hom}}

\revs{In this section we explain connections between condensation and stationary currents for models with bounded and unbounded rates, review previous results on the equivalence of ensembles based on work by one of the authors \cite{stefan}, and state one of our main new results on strong equivalence for subcritical systems.\revf}

\subsection{General remarks\label{sec:genrem}}

For a spatially homogeneous system under the grand-canonical measures the occupation numbers $\eta_x$ are i.i.d. random variables taking values in $\N =\{ 0,1,\ldots\}$, each with distribution 
\bea
\nu_\phi^1 [\eta_x =n]=\frac{1}{z(\phi )}\, w(n)\,\phi^n\quad\mbox{with}\quad z(\phi )=\sum_{n=0}^\infty w(n)\,\phi^n \ ,
\eea
for all $x=1,2,\ldots$. Connecting to our previous notation we have $\lambda_x \equiv 1$ and therefore $\phi_c =\phi_c^x =1$ with (\ref{lambound2}).
We have simply $R_x (\phi )=R(\phi )$ for all $x\in\Lambda$ and the critical density (\ref{rhoc}) is given by
\bea
\rho_c =R(1) =\frac{1}{z(1)}\sum_{n=0}^\infty n\, w(n)\in (0,\infty ]\ .
\eea
It can be shown that $z(1)=\infty$ implies $\rho_c =\infty$ (see e.g. \cite[Lemma II.3.3]{kipnislandim}). Therefore, the system exhibits condensation with $\rho_c <\infty$ if and only if $n w(n)$ is summable, i.e. $w(n)$ has to decay fast enough like a power law or another sub-exponential distribution. In that case the measures are defined for all $\phi\in [0,1] =D_\phi$ and the range of densities is given by $R(D_\phi )=[0,\rho_c ]$.

So for $\rho_c <\infty$ the range of densities attainable by grand-canonical measures is a strict subset of $[0,\infty )$. For typical stationary configurations under the canonical distribution $\pi_{\Lambda ,N}$ with $N/L=\rho >\rho_c$ the system phase separates into a condensed and a fluid phase. As we will review in the following subsections, it can be shown with a general thermodynamic approach and simple relative entropy techniques that the bulk phase is distributed as the product measure $\nu_1$ at the critical density $\rho_c$, and that the condensed phase concentrates on a vanishing fraction of the lattice containing a macroscopic amount of order $(\rho -\rho_c )L$ of particles. This is in general analogous to classical results on phase separation in the Ising model with spin-exchange (Kawasaki) dynamics (see e.g. \cite{liggett}, Chapter 4), the main difference being that the local state space of our models is unbounded and the condensed phase contributes only subextensively to the total free energy (or entropy) of the system. Therefore various classical results on weak convergence of local observables have to be improved and we will discuss this in detail in the following subsections which include new results in this direction. For the special cases of $w(n)$ having power law or stretched exponential tails it has been shown that the condensed phase consists in fact of a single site in a series of papers \cite{jeonetal00,ferrarietal07,armendarizetal09} \revs{and \cite{stefan,agl} involving one of the authors\revf}. This information is not accessible by our thermodynamic treatment, which in principle can also be applied to more general models with non-product stationary measures, that can exhibit a non-trivial structure of the condensed phase (see \cite{evansetal06} and \cite{waclawetal09} and references therein).

For zero-range processes, the weights in fact uniquely determine the dynamics (cf. Section 2.1) via $u(n)=w(n-1)/w(n)$, and a standard example is given by
\bea\label{zrconex}
u(n)=1+\frac{b}{n^\gamma}\quad\mbox{for all }n\geq 1\quad\mbox{and }u(0)=0\ ,
\eea
which has first been studied in \cite{evans00}. The parameters are non-negative, and if $\gamma\in (0,1)$ or $\gamma =1$ and $b>2$ the weights show a stretched exponential or power law decay, respectively, which leads to $\rho_c <\infty$ (see e.g. \cite{evans00,stefan} for details). Heuristically, the dynamic mechanism of condensation in these models is an effective attraction between particles on sites with high occupation numbers, resulting from the asymptotic decay of the jump rates $u(n)$. As a result large clusters of particles become essentially immobile, and receive roughly as many particles from their neighbouring sites as they eject leading to a current balance between condensed and fluid phase. In general lattice gases, the stationary current is defined as the expected net number of particles crossing a bond in a (specified) positive direction per unit of time. The full current depends strongly on the lattice geometry and vanishes for reversible systems, in which case one has to consider the diffusivity. The crucial quantity for our interest is the interaction part of both quantities which is given by the average jump rate of a particle per connecting bond. We will simply call this the current in the following for ease of presentation, having in mind totally asymmetric nearest neighbour systems in one dimension as typical examples. Note that under condition (\ref{homcon}) the dynamics in fact fulfill the gradient condition (see e.g. \cite{spohnbook}, Section II.2.4) which leads to a simplified expression for the diffusivity justifying this simplification also for reversible systems.

Since $\nu_\phi^\Lambda$ is a homogeneous product measure, the grand-canonical current can be defined for an arbitrary pair of sites $x\neq y\in\Lambda$ and is given by
\bea\label{current}
j_{gc} :=\nu_\phi^\Lambda \big( u(\eta_x )\, v(\eta_y )\big) =\nu_\phi^1 (u)\,\nu_\phi^1 (v)=\phi\,\nu_\phi^1 (v)^2\ ,
\eea
where for the last representation we have used the recursive property (\ref{recurs}) of the stationary measures.
Similarly, we define the canonical current
\bea\label{cancurr}
j_{\Lambda ,N} := \pi_{\Lambda ,N} \big( u(\eta_x )\, v(\eta_y )\big)
\eea
which does not factorize, but is still independent of the actual choice of $x\neq y\in\Lambda$ since for homogeneous systems the canonical measures are permutation invariant. The thermodynamic limit of the canonical current
\bea\label{fundia}
j(\rho )=\lim_{L,N\to\infty} j_{\Lambda ,N}\quad\mbox{for all }\rho\geq 0\ ,
\eea
is also called current-density relation or the fundamental diagram of the process. To compare both currents it is often convenient to also view the grand-canonical one as a function of the density using the one-to-one relation $\rho =R(\phi )$ in (\ref{current}), and in this case we write $j_{gc} (\rho )$ which exists only for densities in $[0,\rho_c ]$. 

\begin{figure}
\includegraphics[width=0.49\textwidth]{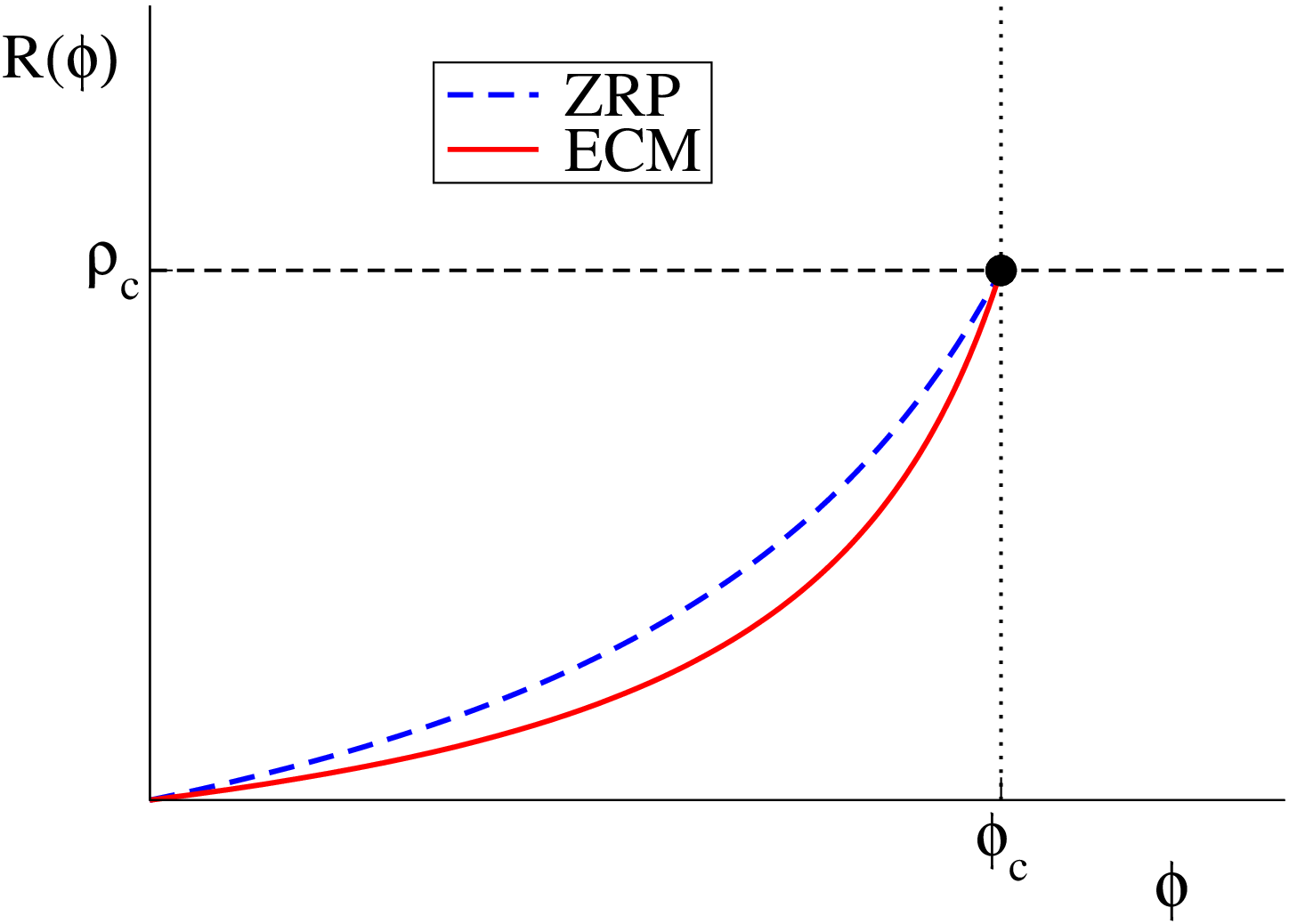}
\hspace{1ex}
\includegraphics[width=0.49\textwidth]{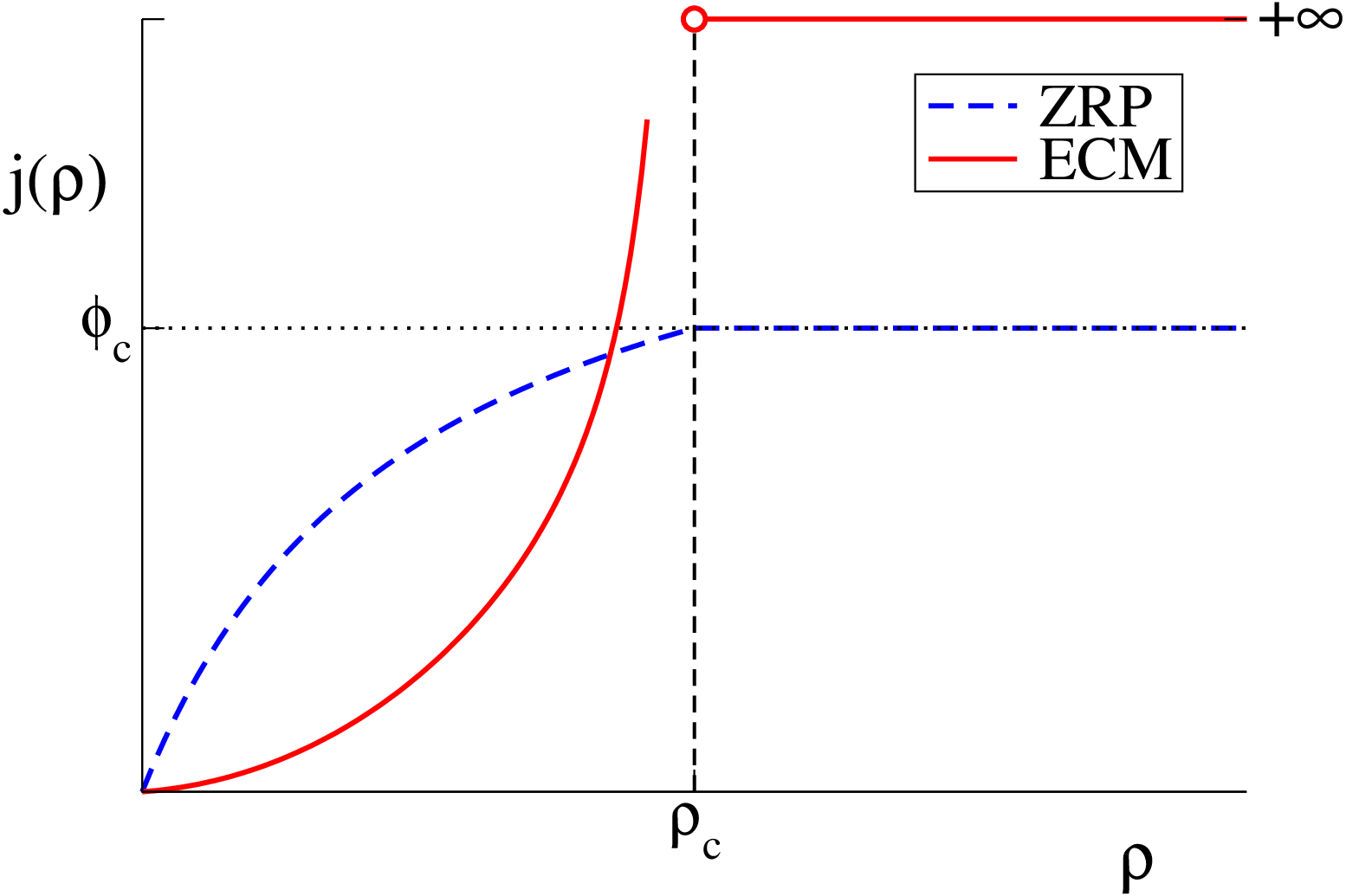}
\caption{\label{fig2}
Sketch of the density $R(\phi )$ (left) and the fundamental diagram $j(\rho )$ (right) for the zero-range process (blue-dashed) and the explosive condensation model (red-solid). While stationary densities (\ref{limdens}) look similar, the systems show very different dynamics and current-density relations (\ref{fundia}), where the mobility of the condensed phase vanishes (ZRP) or diverges (ECM).
}
\end{figure}

For zero-range processes with $v\equiv 1$ (\ref{current}) implies simply $j_{gc} =\phi$, and 
therefore $j_{gc} (\rho )$ is given by the inverse of the function $R(\phi )$, which is illustrated in Figure \ref{fig2}. Since the rates (\ref{zrconex}) are bounded functions, convergence results in the next subsections apply also for supercritical densities so that the fundamental diagram is equal to $j_{gc} (\rho )=j_{gc} (\rho_c )=1$ for all $\rho\geq\rho_c$. This is consistent with the heuristics that the condensed phase in zero-range models is static. From the point of view of conservation laws in the hydrodynamic limit \cite{kipnislandim} this means that the characteristic or group velocity $j'(\rho )$ vanishes for $\rho >\rho_c$, which is consistent with heuristic results in \cite{rosy}.

Another interesting example is given by the explosive condensation model (ECM), recently introduced in \cite{waclawetal11}, with rates given by
\bea
u(n)=(d+n)^\gamma -d^\gamma \ ,\quad v(n)=(d+n)^\gamma \quad\mbox{with }d,\gamma >0\ .
\eea
The stationary weights for this model have leading order asymptotic decay
\bea
w(n)=\prod_{k=1}^n \frac{(k-1+d)^\gamma}{(k+d)^\gamma -d^\gamma} \sim n^{-\gamma}
\eea
for all $d>0$, so the system exhibits condensation for $\gamma >2$. The function $R(\phi )$ is also shown in Figure \ref{fig2} and looks very similar to the ZRP as do typical stationary configurations. However, 
the grand-canonical current (\ref{current}) in the ECM can be written as
\bea
j_{gc} = \phi \big(\nu_\phi^1 (v)\big)^2 =\frac{\phi}{z(1)^2} \Big(\sum_{n\geq 0}w(n)v(n)\phi^n\Big)^2 \to\infty\quad\mbox{as }\phi\nearrow 1\ ,
\eea
since $w(n)v(n)\sim n^{-\gamma} (d+n)^\gamma =O(1)$ as $n\to\infty$. As opposed to the ZRP with rates (\ref{zrconex}) discussed above, the rates are now unbounded functions and their expectation is in fact a higher order moment with power $\gamma$ which diverges as $\rho\nearrow\rho_c$. Intuitively, this leads to a very high mobility of large clusters when the critical density is approached, which in fact diverges in the thermodynamic limit. For supercritical densities the stationary current is dominated by the condensate contribution and diverges as $L^{\gamma -1}$. In the totally asymmetric one-dimensional system studied in \cite{waclawetal11} the condensate will move ballistically across the lattice with diverging speed. This effect is strong enough that the equilibration time of supercritical systems in fact vanishes in the limit $L\to\infty$ (hence the name `explosive condensation'). The dynamics of the model is clearly not well defined in the thermodynamic limit at least for supercritical densities, while this is expected to be the case for the ZRP with bounded rates (\ref{zrconex}), even though this is not proven to our knowledge.

Note that in contrast to the strikingly different dynamics, which is encoded in the different asymptotic behaviour of the jump rates, the static stationary behaviour for both models is in fact identical. Further details on recent rigorous results on the dynamics of condensation will be given in Section \ref{sec:dynamics}, in the following we focus on a detailed study of the equivalence of canonical and grand-canonical measures.


\subsection{General results\label{sec:genres}}

In the following we present results on the equivalence of ensembles between canonical and grand-canonical measures which hold under very general conditions. Building on the following simple theorem for relative entropy densities published previously in \cite{stefan} \revs{involving one of the authors\revf}, we discuss general consequences for the convergence of observables or test functions, and present new results on how they can be extended in sub- and supercritical cases.

\bp{relen}
As $L,N\to\infty$ such that $N/L\to\rho$ we have
\bea
\frac1L H(\pi_{\Lambda ,N};\nu_\phi^\Lambda ) \to 0\ ,
\eea
provided that $\phi\in [0,1]$ is chosen such that $R(\phi )=\rho$ for $\rho <\rho_c$ or $\phi =\phi_c =1$ for $\rho\geq\rho_c$.\\
For $\rho \leq\rho_c$, this also holds with a prefactor $1/a_L$ for any $a_L \gg\log L$, and for $\rho >\rho_c$ we need $a_L \gg\log L\vee t_L$, where $t_L :=-\log\bar\nu_1^1 (L)$ encodes the sub-exponential tail of the critical marginals $\nu_1^1$. 
\ep

\begin{proof} We use the representation (\ref{ldcon})
\bea
\frac1L H(\pi_{\Lambda ,N};\nu_\phi^\Lambda )=-\frac1L \log\nu_\phi^\Lambda \big[\Sigma_\Lambda =N\big]\ ,
\eea
and for $\rho <\rho_c$ the marginals $\nu_\phi^1$ with $R(\phi )=\rho$ have exponential tails and the standard local limit theorem \cite{mcdonald1} (see also appendix) provides an upper bound of order $\log L/L$. For $\rho =\rho_c$ the sub-exponential tails of $\nu_1^1$ can lead to diverging second moments, but in any case the local limit theorem for non-normal limit distributions (see e.g. \cite{mitalauskas}) provides a lower bound of order $\nu_\phi^\Lambda \big[\Sigma_\Lambda =N\big]\geq 1/L$ which leads to the same conclusion. 
For $\rho >\rho_c$ this is a large deviation probability, and a simple lower bound is given by putting all excess mass in the first site,
\bea
\nu_1^\Lambda \big[\Sigma_\Lambda =N\big]\geq \bar\nu_1^1 \big( N-[\rho_c L]\big)\,\nu_1^{\Lambda\setminus 1} \big[\Sigma_{\Lambda\setminus 1} =[\rho_c L]\big]\ .
\eea
Since $\bar\nu_1^1 (n)=w(n)$ it has sub-exponential tails in the strong form (\ref{subass}) which implies that $\frac1L\log\bar\nu_1^1 \big( N-[\rho_c L]\big)$ vanishes with a speed $t_L /L$ depending on the actual tail, and the second term vanishes analogously to the above.
\end{proof}

Following a classical result of Pinsker (\cite{pinsker}, or \cite[Lemma 5.2.8]{gray}), relative entropy provides an upper bound for the total variation distance $d_{TV}$ of two measures $\mu_1 \ll\mu_2$. This can be written as (see e.g. \cite{peresbook}, Section 4.1)
\bea
d_{TV} (\mu_1 ,\mu_2) =\frac12\sum_{\omega\in\Omega} \big| h(\omega ) -1\big|\, \mu_2 (\omega)=\frac12\mu_2 \big( |h-1|\big)\ ,
\eea
where $h$ is the Radon-Nikodym derivative.

\bl[(Pinsker)]{pinsker}
The total variational distance of two measures $\mu_1 ,\mu_2$ is bounded above by the relative entropy as
\bea
d_{TV} (\mu_1 ,\mu_2)\leq \sqrt{2H(\mu_1 ;\mu_2)}\ .
\eea
\el

\noindent We use this together with sub-additivity of relative entropy \cite{csiszarkoerner} to formulate general implications of the above theorem on convergence of local test functions.
Since $\pi_{\Lambda ,N}\ll\nu_\phi^\Lambda$ the Radon-Nikodym derivative $h_\Lambda =\pi_{\Lambda ,N} /\nu_\phi^\Lambda$ exists, and since $\pi_{\Lambda ,N} =\nu_\phi^\Lambda [.|\Sigma_\Lambda =N]$ it is given by
\bea
h_\Lambda (\eta )=\frac{1}{\nu_\phi^\Lambda [\Sigma_\Lambda =N]}\,\1_{\Sigma_\Lambda =N} (\eta )\ .
\eea
The derivative of the marginal distributions on a subset $\Delta\subset\Lambda$ is given by
\bea\label{rdmar}
h_\Lambda^\Delta (\eta^\Delta )=\frac{\nu_\phi^{\Lambda\setminus\Delta} \big[\Sigma_{\Lambda\setminus\Delta} =N-|\eta^\Delta |\big]}{\nu_\phi^\Lambda [\Sigma_\Lambda =N]}\,\1_{\Sigma_\Delta \leq N} (\eta^\Delta )\ .
\eea

\bc{gencon}
As $L,N\to\infty$ such that $N/L\to\rho$ we have for all finite $\Delta\subset\N$
\bea\label{l1con}
d_{TV} \big( \pi_{\Lambda ,N}^\Delta ,\nu_\phi^\Delta \big)=\nu_\phi^\Delta \big( |h_\Lambda^\Delta -1|\big)\to 0\ ,
\eea
provided that $\phi\in [0,1]$ is chosen such that $R(\phi )=\rho$ for $\rho <\rho_c$ or $\phi =\phi_c =1$ for $\rho\geq\rho_c$. This is equivalent to local weak convergence, i.e. for all bounded cylinder functions $f\in C_0^b (X)$ we have
\bea
\pi_{\Lambda ,N} (f)\to \nu_\phi (f)\ .
\eea
\ec

\begin{proof}
By subadditivity of relative entropy (see e.g. \cite{csiszar84}) we get from Proposition \ref{relen}
\bea
H(\pi^\Delta_{\Lambda ,N};\nu_\phi^\Delta )\leq C\frac{|\Delta |}{L} H(\pi_{\Lambda ,N};\nu_\phi^\Lambda )\to 0\quad\mbox{as }L\to\infty\ ,
\eea
and the first claim follows immediately from Pinsker's inequality.
For $f\in C_0^b (X)$ we pick $\Delta$ large enough to include its support and get
\bea
\big| \pi_{\Lambda ,N} (f)-\nu_\phi (f)\big|\leq \nu_\phi^\Delta \big( |f(h_\Lambda^\Delta -1)|\big) \leq\| f\|_\infty \nu_\phi^\Delta \big( |h_\Lambda^\Delta -1|\big)\to 0
\eea
as $L\to\infty$, which implies the second statement since $f$ is bounded.
\end{proof}

\revs{This result was previously published involving one of the authors in \cite{stefan}.\revf} 
By general compactness arguments on the limiting statespace $X=\N^\N$ (which is itself non-compact) presented e.g. in \cite[Lemma II.1.2]{kipnislandim}, convergence of bounded cylinder test functions implies (global) weak convergence, i.e. convergence of expectations of all bounded functions $f\in C^b (X)$. To formulate this precisely, one has to extend the definition of the canonical measures to the limiting state space on the infinite lattice, which is usually done by periodic extensions. In a similar fashion, explicit upper bounds on the relative entropy density can be used to derive total variation convergence of marginals on subextensive volumes $\Delta$ with $|\Delta |/L\to 0$ fast enough.

Corollary \ref{gencon} contains classical implications of relative entropy convergence which have been derived for spin systems and are satisfactory in this context, but in our case of systems with non-compact local state space convergence of bounded test functions is a very weak statement. In fact, not even the density on a site (given by $f(\eta )=\eta_x)$ is bounded, and in general also does not converge due to the condensation phenomenon. It is therefore desirable to strengthen the above result which is discussed in the following.

\subsection{Subcritical systems\label{sec:subcrit}}

The main \revs{new\revf} result of this section gives a strong version of weak convergence for integrable test functions, using extra regularity of the Radon-Nikodym derivatives $h$ (\ref{rdmar}) in subcritical systems. It provides an elegant extension of a result in \cite[Appendix 2]{kipnislandim}, for $L^2$-functions with a rather complicated proof involving the Cramer expansion.

\bt{equi}
For $\rho\leq\rho_c$ and $R(\phi )=\rho$ we have $\limsup_{L\to\infty} \| h_\Lambda^\Delta \|_\infty <\infty$ for every finite subset $\Delta\subset\Lambda$. Furthermore,
\bea
\big| \pi_{\Lambda ,N} (f)-\nu_\phi (f)\big|\to 0\quad\mbox{for all }f\in L^{1+\epsilon}(\nu_\phi )\cap C_0 (X)
\eea
for some $\epsilon >0$ in the thermodynamic limit.
\et

\begin{proof}
By the local limit theorem the supremum of $h_\Lambda^\Delta$ for large enough $L$ is obtained for $|\eta^\Delta |\sim \rho |\Delta |$ and we have with an $L$-independent constant $C_\rho$
\bea
\| h_\Lambda^\Delta\|_\infty \leq C_\rho \frac{\nu_\phi^{\Lambda\setminus\Delta} [\Sigma_{\Lambda\setminus\Delta} =N-\rho |\Delta |]}{\nu_\phi^\Lambda [\Sigma_\Lambda =N]}\ .
\eea
Again by the local limit theorem both terms are of the same order in $L$, and for convergence to a Gaussian law we have for all $L$ large enough
\bea
\| h_\Lambda^\Delta\|_\infty \leq C'_\rho \frac{1/\sqrt{L-|\Delta |}}{1/\sqrt{L}} =\frac{C'_\rho}{\sqrt{1-|\Delta |/L}} <\infty\ .
\eea
For convergence to other stable laws an analogous estimate holds with powers different from $1/2$ for which the argument still works.\\
Now fix $\epsilon >0$ and a cylinder function $f\in L^{1+\epsilon} (\nu_\phi )$. 
Then picking $\Delta$ large enough to contain the support, we can use H\"older's inequality to get
\bea
\big| \pi_{\Lambda ,N} (f)-\nu_\phi (f)\big|\leq C\big\| f(h_\Lambda^\Delta -1)\big\|_1 \leq C\| f\|_{1+\epsilon} \| h_\Lambda^\Delta -1\|_{1+1/\epsilon}\ ,
\eea
where norms are w.r.t. the measure $\nu_\phi$. This bound vanishes as $L\to\infty$, since
\bea
\| h_\Lambda^\Delta -1\|_{1+1/\epsilon} \leq \Big(\big( 1+\| h_\Lambda^\Delta \|_\infty \big) \| h_\Lambda^\Delta -1\|_1 \Big)^{\epsilon /(1+\epsilon)} \to 0\ ,
\eea
which finishes the proof.
\end{proof}

For subcritical systems with $\rho <\rho_c$ this result includes in particular convergence for all polynomial moments and also some exponential moments. Furthermore, in the proof of the first statement it suffices to take $\Delta\subset\Lambda$ such that $\limsup_{L\to\infty} |\Delta |/L <1$. Thus, as long as one measures only on a fraction of the volume the canonical measure is conditioned on, it is asymptotically equivalent to the product measure with respect to a much larger class of $L^{1+\epsilon}$ integrable test functions for any $\epsilon >0$. Again we do not state this explicitly to avoid the technical issue of extending the canonical measures which does not provide much insight.

\subsection{Supercritical systems\label{sec:supcrit}}

For supercritical, phase separated systems we cannot expect to improve the general results of Corollary \ref{gencon}, unless we restrict attention to the fluid phase. This may seem simple since the condensed phase only covers a vanishing volume fraction and is delocalized in the thermodynamic limit, however, it carries a finite fraction of the mass and therefore contributes when measuring the particle density or higher moments. Since the contribution of the condensed phase to averages concentrates on high occupation numbers which diverge in the limit, the simplest way to restrict to the fluid phase is to consider a sequence of bounded functions via cut-off, for which the general weak convergence results can be directly applied.

\bc{fluidtest}
For any integrable cylinder test function $f\in L^1 (\nu_1 )\cap C_0 (X)$ we have in the thermodynamic limit with $N/L\to\rho >\rho_c$
\bea\label{klim}
\lim_{K\to\infty}\lim_{L\to\infty}\pi_{\Lambda ,N} \big( f\wedge K\big) =\nu_1 (f)\ .
\eea
\ec

\begin{proof}
We use the obvious upper bound of $\big|\pi_{\Lambda ,N} \big( f\wedge K\big) - \nu_1 (f)\big|$
\bea\label{estima}
\big|\pi_{\Lambda ,N} \big( f\wedge K\big)-\nu_1 (f\wedge K)\big| +\big|\nu_1 (f\wedge K)-\nu_1 (f)\big|\ ,
\eea
and the first part vanishes due to weak convergence (Corollary \ref{gencon}) in the limit $L\to\infty$ for every fixed $K$. The second $L$-independent part then vanishes as $K\to\infty$ by dominated convergence.
\end{proof}

If $f$ itself is unbounded the limits do in general not commute, for example $\pi_{\Lambda ,N} (\eta_x )\to\rho$, whereas $\pi_{\Lambda ,N} (\eta_x \wedge K)\to\rho_c$ for all $K\geq 0$. From the results in Section \ref{sec:genres} we have explicit bounds on the first term in (\ref{estima}), 
\bea\label{tailcond}
|..|\leq K\sqrt{\frac{\log L\vee t_L}{L}}\quad\mbox{where}\quad t_L =-\log\bar\nu_1^1 (L)\ .
\eea
So certain joint limits with $K=K_L$ are possible in (\ref{klim}) depending on the sub-exponential tail of the critical measure encoded in $t_L \ll L$.

Another more involved approach is to `localize' the condensed phase which also gives additional information about its spatial extension. Since the grand-canonical measures are of product form, it turns out that the condensed phase in fact consists of a single lattice site and can therefore be identified with the maximum. This has been established in \cite{armendarizetal09} for power-law and stretched exponential tails of $\nu_1^1$, with complementing results also in \cite{jeonetal00,agl}. The proof requires explicit estimates and we only quote the main result here in a slight reformulation.

Since the canonical measures are permuation invariant, we can just consider conditional measures to localize the condensed phase
\bea
\tilde\pi_{\Lambda ,N} =\pi_{\Lambda ,N} \big[ .\big|\,\eta_1 =M_{\revs{\Lambda\revf}} \big]\quad\mbox{where}\quad M_\Lambda (\eta )=\max_{x\in\Lambda} \eta_x \ .
\eea

\bt[\cite{armendarizetal09}]{al}
Assume that $\nu_1^1$ has a power-law tail or a stretched exponential tail with $\rho_c <\infty$. In the thermodynamic limit $L,N\to\infty$ with $N/L\to\rho >\rho_c$
\bea
d_{TV} \big( \tilde\pi_{\Lambda ,N}^{\Lambda\setminus 1} ,\nu_1^{\Lambda\setminus 1}\big)\to 0\ .
\eea
\et

This is significantly stronger than the general local result in Corollary \ref{gencon}, stating that in the limit all but the maximum behave as i.i.d. random variables with distribution $\nu_1^1$ with convergence in total variation distance. This implies in particular a law of large numbers and a central limit theorem for the occupation of the maximum, which contains on average all the excess mass $(\rho -\rho_c )L$. Fluctuations are Gaussian on scale $\sqrt{L}$ if $\nu_1^1$ has finite variance, and otherwise obey standard stable law fluctuations on larger scales (see Corollary 1 in \cite{armendarizetal09}). Also higher order statistics are included, for example the largest component in the bulk obeys standard extreme value statistics for i.i.d. random variables under $\nu_1^1$.

Note that equivalence in the bulk for condensed systems is even stronger than for subcritical systems in Section \ref{sec:subcrit}, where equivalence to product measures holds at most on finite volume fractions. This is due to the fact that the number of particles in the canonical measures is fixed, and the fluctuations in the observation volume have to be compensated by the rest of the system, which has to be big enough to achieve this for typical configurations. In the supercritical case, all the fluctuations of the bulk can be compensated by the condensate on a single site since it contains an extensive number of particles.

\section{Condensation in inhomogeneous systems\label{sec:inhom}}

In this section we review previous work on finite inhomogeneous systems and present new results on the equivalence of ensembles in the thermodynamic limit for sub- and supercritical systems in Sections 4.3 and 4.4.

\subsection{General remarks}

\begin{figure}

\begin{center}
\includegraphics[width=0.65\textwidth]{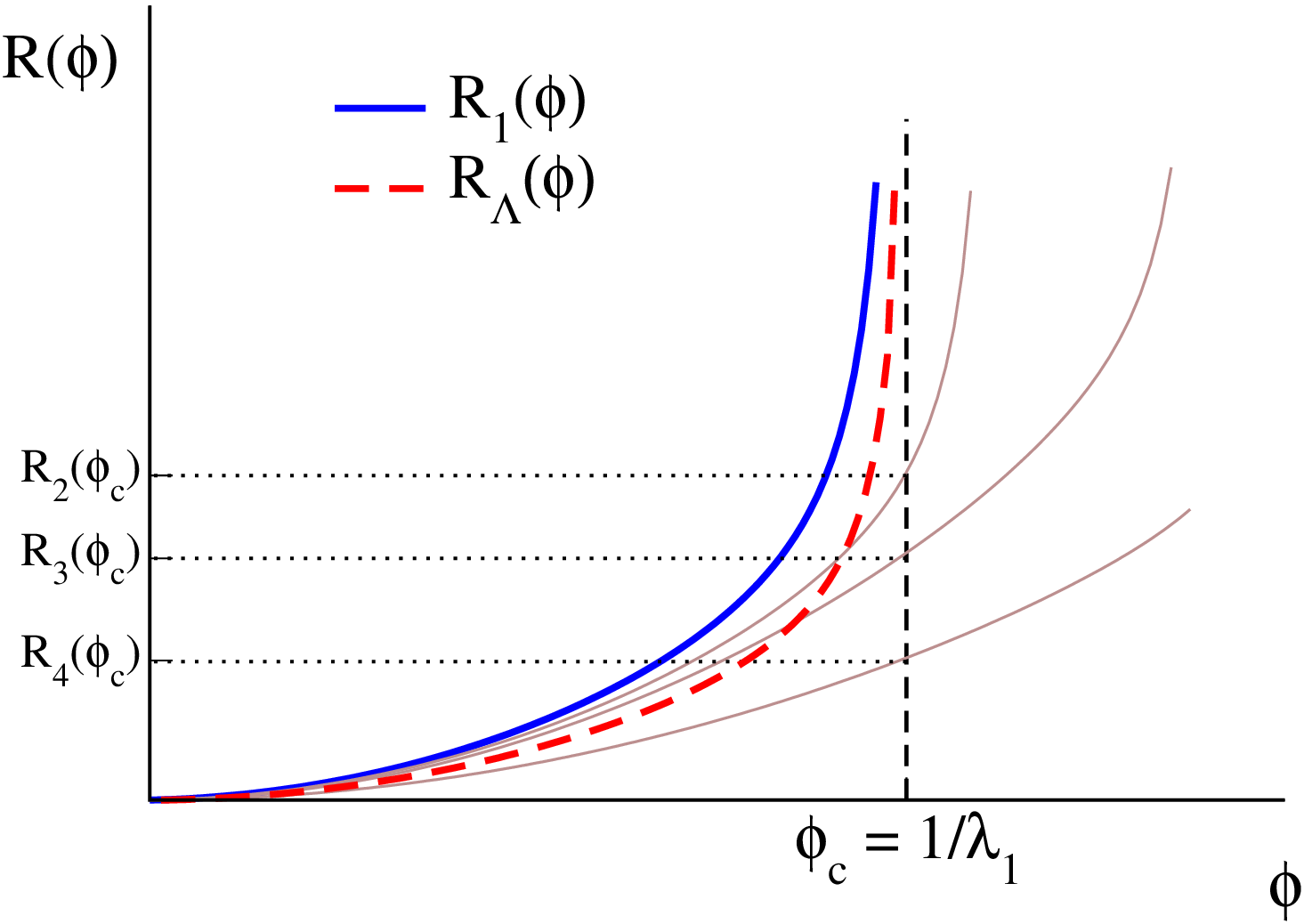}
\end{center}

\caption{\label{figfinite}
Condensation in finite inhomogeneous systems with a trap site at $1$ and $\phi_c =1/\lambda_1$. For $\phi\nearrow\phi_c$ the density $R_1 (\phi )$ diverges, forming the condensate. All other sites are asymptotically distributed as the critical product measure $\nu_{\phi_c}^{\Lambda\setminus\Delta}$ with densities $R_x (\phi_c )<\infty$, forming the fluid phase (examples for $x=2,3,4$ shown in grey). The diverging condensate contribution dominates the system density $R_\Lambda (\phi_c )$ (\ref{decomp}) and the domain is $D_\phi =[0,\phi_c )$.
}
\end{figure}

For spatially inhomogeneous systems condensation can also be caused by the presence of trap sites, which are characterized by large values of $\lambda_x$ and therefore a slow decay of the tail. Recall that the marginals have the form
\bea
\nu_\phi^x [\eta_x =n]=\frac{1}{z_x (\phi )}w_x (n)\, (\phi\lambda_x )^n
\eea
with sub-exponential weights $w_x (n)$. 
The phenomenon is easiest explained on a fixed lattice $\Lambda$, where we have
\bea
\phi_c =\min_{x\in\Lambda} 1/\lambda_x \ .
\eea
Assume that $\phi_c =1/\lambda_y$ for all $y\in\Delta$ for some subset of trap sites $\Delta\subsetneq\Lambda$. Note that even for $\Delta =\Lambda$ there could still be condensation due to the interaction mechanism in $w_x (n)$ as described for homogeneous systems above. We do not consider this here and will comment on it later. As a simple example one can think of $\Delta =\{ 1\}$ as being a single site, which is the case in the illustration in Figure \ref{figfinite}. For fixed lattices $\Lambda$ the condensate will dominate the fluid phase and the characterization of condensation in Definition \ref{condensation} has to be adapted in an obvious way. The contribution to the total density in the system can be divided as 
\bea\label{decomp}
R_\Lambda (\phi )=\frac1L\sum_{x\in\Delta} R_x (\phi )+\frac1L\sum_{x\in\Lambda\setminus\Delta} R_x (\phi )\ .
\eea
As $\phi\nearrow\phi_c$ the second contribution converges and the first one diverges, corresponding to a phase separation into condensed sites $\Delta$ and bulk sites $\Lambda\setminus\Delta$. So when conditioned on a very high particle number $N$, most of the mass will concentrate on the condensed sites $\Delta$ and sites in the fluid phase will be distributed according to the critical product measure $\nu_{\phi_c}^{\Lambda\setminus\Delta}$ with densities $ R_x (\phi_c )<\infty$.

Note that for fixed $\Lambda$, $R_\Lambda$ is unbounded and (\ref{decomp}) can be solved for $\phi_N$ such that $R_\Lambda (\phi_N )=N/L$ for any $N$. This is a crucial difference to homogeneous systems where $R_\Lambda =R_1$ is independent of the system size and bounded above by $\rho_c$ for all $\Lambda$. In the limit $N\to\infty$ this leads to a sequence $\phi_N \nearrow \phi_c$ describing the exact distribution of mass between condensate and fluid phase. This limit for fixed $\Lambda$ has been \revs{derived by direct computation in \cite{getal11} involving one of the authors,\revf} and we just quote the result without proof.

\bp[\cite{getal11}]{fininhom}
For fixed $\Lambda$ the bulk marginal converges weakly to the critical product measure, i.e. for all bounded, continuous $f\in C^b (X)$ 
\be\label{finihom1}
\pi_{\Lambda ,N}^{\Lambda\setminus\Delta} (f)\to \nu_{\phi_c}^{\Lambda\setminus\Delta} (f)\quad\mbox{as }N\to\infty\ .
\ee
The condensed phase contains almost all particles, i.e. for all $\delta\in (0,1)$
\be\label{finihom2}
\pi_{\Lambda ,N} \big[ \Sigma_\Delta \geq\delta N\big] \to 1\quad\mbox{as }N\to\infty\ .
\ee
Furthermore, we have a strong law of large numbers where $\Sigma_\Delta /N\to 1$ almost surely.
\ep

The case $\lambda_x =\infty$ for some $x\not\in\Delta$ can be included as well, and in those sites $R_x (\phi )<\infty$ is defined for all $\phi\geq 0$. Note that the distribution of mass in the condensed phase depends on the asymptotic behaviour of the weights $w_x (n)$ for $x\in\Delta$ and cannot be further specified in the general case. The case $\lambda_x \equiv 1$ with spatial disorder only in the sub-exponential weights $w_x (n)$ studied in \cite{luis1,luis2} leads to a rather complicated behaviour and very slow convergence of critical observables in the limit of large system sizes. More recently, the interplay between spatial disorder and sub-exponential tails has been studied in \cite{godrecheetal12}, leading to a rich phase diagram where condensation can be dominated by either one or a combination of the interaction or the spatial mechanism.

In all of this section we focus on condensation originating from spatial inhomogeneity of the $\lambda_x$, which is the case as soon as $\lambda_x$ take at least two different values. This regime has been studied before \cite{evans96,krugetal96,benjaminietal96} for a special class of spatially inhomogeneous zero-range processes with rates $u_x (n)\equiv u_x$ for $n\geq 1$, corresponding to $w_x (n)\equiv 1$ for all $n\geq 0$. Trap sites in this model are simply the ones with the slowest jump rate $u_x$. Results with more general rates in \cite{landim96,andjeletal00,ferrarisisko} still require rates $u_x (n)$ to increase monotonically with $n$, and make crucial use of the resulting attractivity of the model and coupling techniques, providing also dynamical results. Our contribution in the next subsections generalizes this to much more general product measures with mild assumptions on regularity. 
This approach is based on the thermodynamic methods introduced previously which do not require any assumptions on monotonicity, at the cost of providing less detailed statements.
Further references related to hydrodynamic limits of disordered lattice gases are provided in the introduction of \cite{ferrarisisko}.

\subsection{The thermodynamic limit}

In the thermodynamic limit $\Lambda\nearrow\N$ we have
\bea\label{domlim}
\phi_c =\inf_{x\in\N} 1/\lambda_x
\eea
and the infimum is not necessarily attained on any site $x$. In the following we use the definition (\ref{limdens}) of $R(\phi )$ for the limiting system density and  (\ref{rhoc}) 
of the critical density. 
\revs{Since the\revf} condensed phase does not dominate the system in the thermodynamic limit,
\revs{we are able to\revf} adopt Definition \ref{condensation} as a general characterization as for homogeneous systems. In addition to the regularity assumptions (\ref{lambound2}) and (\ref{wbound}) on the weights $w_x$, we assume that
\bea\label{werner}
\sum_{n\geq 0} n\, w_-(n)=\infty\ ,
\eea
which implies that $R_x (\phi )\to\infty$ when $\phi\nearrow\phi_c^x$ approaches the radius of convergence of $z_x$. This rules out condensation being caused by particle interactions as in Section \ref{sec:hom}, and will allow for a much more coherent presentation of the phenomenon due to spatial inhomogeneities. As an immediate consequence, $R_\Lambda (\phi )$ diverges as $\phi\nearrow\phi^\Lambda_c$ on finite lattices, and for all densities $\rho >0$ there is a fugacity $\phi_\Lambda (\rho )$ solving $R_\Lambda (\phi )=\rho$. In the thermodynamic limit $R_\Lambda$ can converge pointwise to a bounded function on $D_\phi$, which characterizes condensation. This is a major difference to homogeneous condensation, where $R_\Lambda =R$ is already bounded for all $\Lambda$.
To demonstrate that assumption (\ref{werner}) is not crucial, we will formulate a slightly weaker result at the end of the next subsection without this assumption.

We first consider a few generic examples some of which have been studied previously, and discuss whether they exhibit condensation in the sense of Definition \ref{condensation} which is purely a condition on grand-canonical measures. In the next subsection we give results confirming that this indeed implies phase separation and condensation in the sense of the equivalence of ensembles with canonical measures.
For simplicity we take $w_x (n)\equiv 1$ in the examples, but all general argments hold under the above assumptions.

\begin{figure}
\includegraphics[width=0.49\textwidth]{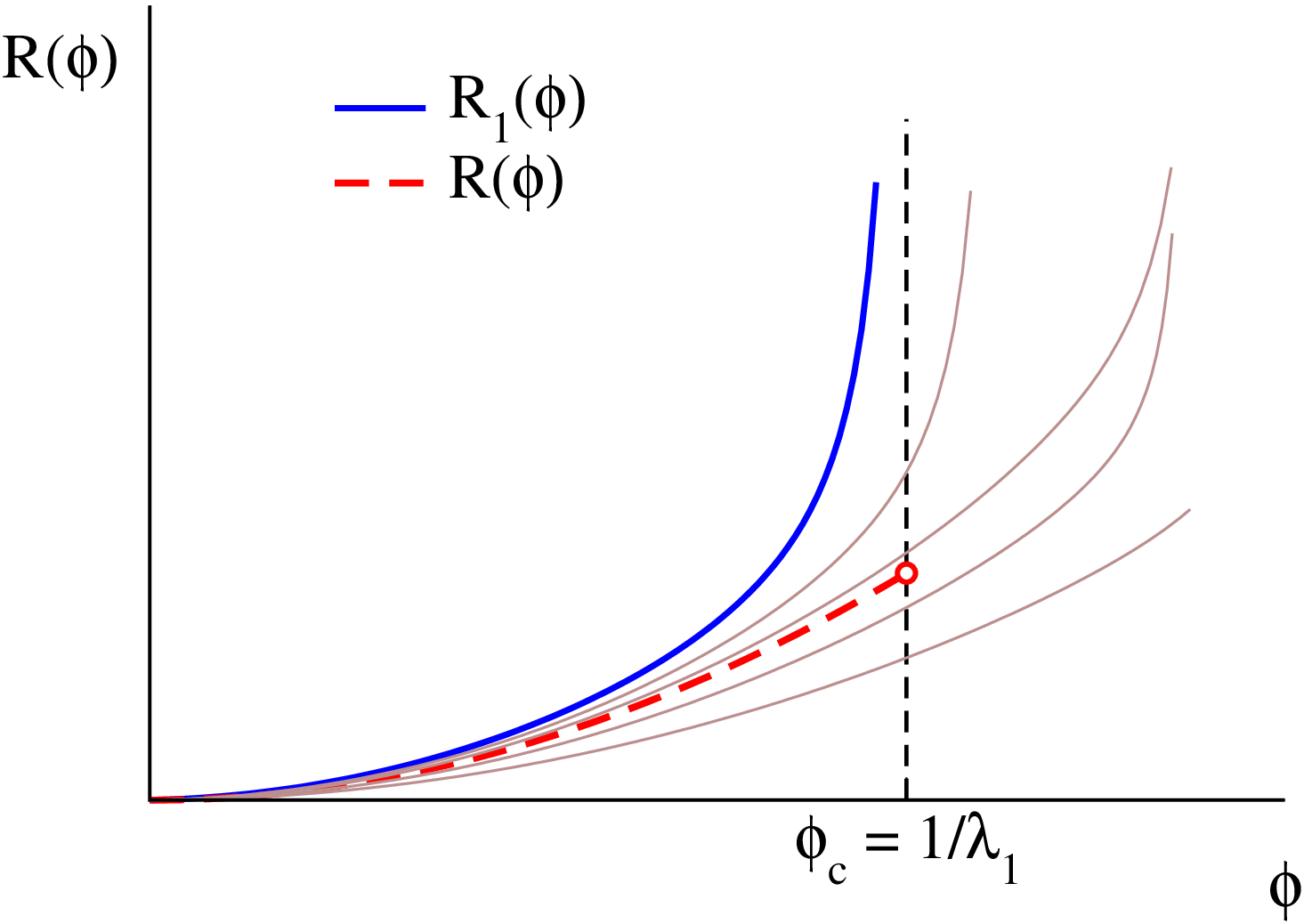}
\hskip1ex
\includegraphics[width=0.49\textwidth]{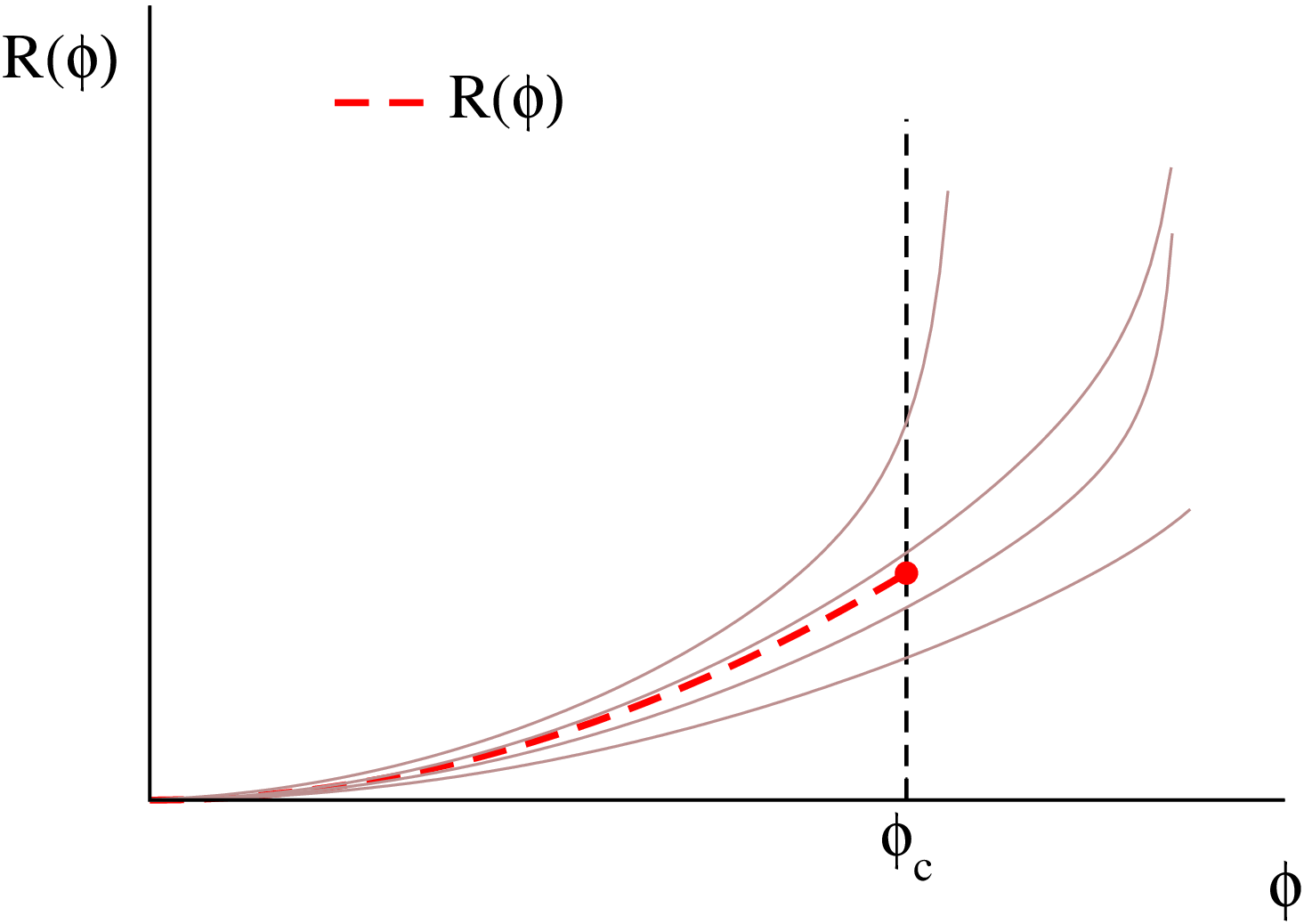}

\caption{\label{fignew}
Sketch of densities $R_x (\phi )$ on individual sites, and average density $R(\phi )$ in the thermodynamic limit. For localized systems (left)  with $\phi_c =1/\lambda_1$ the condensate contribution to $R (\phi_c )$ (\ref{decomp}) diverges and the domain is $D_\phi =[0,\phi_c )$. If the infimum in (\ref{domlim}) determining $\phi_c$ is not attained (right), $D_\phi =[0,\phi_c ]$ and the condensed phase can be delocalized and does not contribute to the limiting average density. As in Fig.~\ref{figfinite}, sample curves $R_x (\phi )$ for $x>1$ are shown in grey.
}
\end{figure}

\vskip1ex
\noindent\textbf{Deterministic profiles.} Let $\lambda_1 ,\lambda_2 ,\ldots$ be a deterministic profile such that $\lambda_x \to 1$ as $x\to\infty$. Such profiles can for example arise in driven processes on a semi-infinite lattice \cite{getal11}. We have
	\bea\label{rex}
	R(\phi )=\lim_{L\to\infty} \frac1L\sum_{x\in\Lambda} R_x (\phi )=\lim_{x\to\infty} R_x (\phi )=\frac{\phi}{1-\phi}\ .
	\eea
	Whether or not the system exhibits condensation then simply depends on the domain $D_\phi$. For an increasing profile $\lambda_x \nearrow$ we have for example $\phi_c =1$, $D_\phi =[0,1)$ and the system does not condense since $\rho_c =\infty$.\\
	If $\lambda_x >1$ for some $x$ a maximum will be attained, we assume it is $\lambda_1 >1$ for simplicity as is for example the case for a decreasing profile $\lambda_x \searrow$. Then $D_\phi =[0,1/\lambda_1 )$ and the system condenses with critical density
	\bea
	\rho_c =\lim_{\phi\nearrow \revs{1/\lambda_1\revf}}R(\phi )=\frac{1}{\lambda_1 -1}\ .
	\eea
	Note that in this case $D_\phi$ is limited only by a single (or in general finite) number of sites, and is strictly smaller than the maximal possible domain $[0,1)$ of the function $R$ (\ref{rex}), as illustrated in Figure \ref{fignew} (\revs{left\revf}). The simplest example of this kind is a single defect site with $\lambda_x >1$, which has been studied in \cite{landim96,angeletal04}. More complicated examples can include arbitrary non-extensive sequences $\lambda_{x_k}$ of defect sites $x_k$, which limit the domain $D_\phi$ via (\ref{domlim}) but do not contribute to the limit in (\ref{rex}). If the subsequence does not attain a maximum, this leads to closed domains of the form $D_\phi =[0,\phi_c ]$ and is illustrated in Figure \ref{fignew} (right).
\vskip1ex
\noindent\textbf{Disordered profiles.} Let $\lambda_1 ,\lambda_2 ,\ldots$ be a sequence of i.i.d. random variables in a compact interval $[0,1/\phi_c ]$, distributed with density $q(\lambda )$. For simplicity assume $\phi_c =1$, then by ergodicity
	\bea
	R(\phi )=\E_q R_x (\phi )=\int_0^{1} \frac{\phi\lambda}{1-\phi\lambda} q(\lambda )d\lambda\ .
	\eea
	Thus, if $q$ is uniformly distributed we see that $R(\phi )$ diverges as $\phi\nearrow\phi_c =1$ and the system does not condense. Intuitively, even though $\lambda_x <1$ for all $x$ with probability one, there are too many sites with $\lambda_x$ very close to $1$, which provide a diverging contribution to the density (\ref{rex}) at $\phi =1$. Condensation is possible if $q$ is small enough near $\lambda =1$, a common choice from previous work \cite{evans96,krugetal96,godrecheetal12} is $q(\lambda )=c(1-\lambda )^{c-1}$. If $c>1$ the system exhibits condensation with critical density
	\bea
	\rho_c =R(1)=\int_0^1 \frac{c\lambda}{1-\lambda} (1-\lambda )^{c-1}d\lambda =\frac{1}{c-1} <\infty\ .
	\eea
	Note that the domain $D_\phi =[0,1]$ contains $\phi_c =1$ in this case. This example can easily be extended to stationary, ergodic environments in the limit $\Lambda\nearrow\Z$ taking values in general intervals $[a,b]$.\\

In general, if the infimum in (\ref{domlim}) is not attained, it has to be approached by a sequence $\lambda_{x_k}$. In this case the condensed phase will be located further and further in the bulk of the system and cannot be measured locally in the limit $L\to\infty$, and we say that it is \textbf{delocalized}. A characterization of this situation in the sense of Definition (\ref{condensation}) is that the domain $D_\phi =[0,\phi_c ]$ is closed, and the critical product measure $\nu_{\phi_c}$ with density $\rho_c =R(1)$ exists in the limit. This is illustrated in Figure \ref{fignew} (right), simple examples are disordered profiles. Note that the sequence $x_k$ necessarily has to be subextensive, i.e. $\big|\{ x_k :k\in\N\}\cap\Lambda \big| /L\to 0$ as $L\to\infty$, otherwise it would provide a diverging contribution to $\rho_c$.

If the infimum in (\ref{domlim}) is attained on a non-empty set of sites $\Delta$, obviously $D_\phi =[0,\phi_c )$. Also $R_x (\phi_c )=\infty$ for all $x\in\Delta$ due to (\ref{werner}) and therefore $|\Delta\cap\Lambda|$
\revs{must again be} subextensive since otherwise this would imply $\rho_c =\infty$ by continuity of $R$. If in addition all other sites have $\lambda$-values uniformly bounded away, i.e. $\lambda_y <1/\phi_c -\delta$ for all $y\not\in\Delta$ and some $\delta >0$, then the condensed phase will be \textbf{localized} in $\Delta$. The excess mass of order $(\rho -\rho_c )L$ will be shared according to details of the weights, but due to the subextensive volume each site will carry a diverging number of particles in the thermodynamic limit. The simplest example is $\Delta =\{ 1\}$ as for decreasing profiles, which is also illustrated in Figure \ref{fignew} (left).

In general, a non-empty set $\Delta$ where the infimum in (\ref{domlim}) is attained and sequences $x_k$ with $\lambda_{x_k} \nearrow 1/\phi_c$ could both exist, and the condensed phase can split into a localized and a delocalized part. The ratio of the excess mass on both parts depends on the details and either of them could also contain the whole condensate. It is also possible that the mass ratio depends on the system size $L$ and does not converge as $L\to\infty$. Constructions of specific profiles $\lambda_x$ and sequences $x_k$ can be attempted along the following lines, tuning the density ratio of a localized defect site e.g. at $x=1$ and the sequence,
\bea\label{ratio}
R_1 (\phi_L )/R_{x_k} (\phi_L )\ .
\eea
Here $\phi_L$ is implicitly determined by
\bea
R_\Lambda (\phi_L )=\sum_{x\in\Lambda} R_x (\phi_L )=\rho\ ,
\eea
which has a unique solution $\phi_L (\rho )$ for each system size $L$ and density $\rho$. If $\rho >\rho_c$, $\phi_L \nearrow\phi_c =1$ (cf. also previous subsection), and there is enough freedom to choose a subextensive sequence $x_k$ and $\lambda_{x_k} \nearrow 1$ to achieve different behaviour in (\ref{ratio}). A priori this leads to rather artificial examples, and it would be interesting to investigate if there are natural situations where a split in a localized and delocalized condensate appears. This is most relevant for disordered profiles where $x_k$ is determined by a record sequence of the $\lambda_{x}$, and has been studied in the context of condensation and growing networks in \cite{godluck08,godluck10}.

\subsection{Equivalence for subcritical systems}

In the following we will show that systems with $\rho_c <\infty$ according to Definition \ref{condensation} indeed exhibit condensation in the sense of the equivalence of ensembles with canonical measures, using an analogous approach as in Section \ref{sec:hom} for homogeneous systems. All results in this section require the regularity assumptions (\ref{wbound}) on the stationary weights and (\ref{lambound2}) on uniform boundedness of the $\lambda_x$. The first \revs{new\revf} result on subcritical systems or systems with $\rho_c =\infty$ can be proved in exactly the same way as for homogeneous systems, envoking a more general version of the local limit theorem \cite{mcdonald1,mcdonald2} which we also summarize in the Appendix.

\bt{subinhom}
For every uniformly bounded sequence $\lambda_1 ,\lambda_2 ,\ldots$ we have 
    \begin{align}
      \frac{1}{L}H\left(\p_{\L,N} ;\nu^{\L}_{\phi}\right)
      \to 0\, \quad \textrm{as $L\to \infty$ and $N/L \to \rho$\ ,}
    \end{align}
provided that $\rho <\rho_c$ (\ref{rhoc}) and $\phi\in D_\phi$ is chosen such that $R(\phi )=\rho$.\\
Furthermore, we have convergence of integrable cylinder functions,
\bea
\big|\pi_{\Lambda,N} (f)-\nu_\phi (f)\big|\to 0\quad\mbox{for all }f\in L^{1+\epsilon} (\nu_\phi )\cap C_0 (X)
\eea
for some $\epsilon >0$ in the thermodynamic limit.\\
Both statements hold also for $\rho =\rho_c$ if the limit measure $\nu_{\phi_c}$ exists and has finite second moments $\nu_{\phi_c} (\eta_x^2 )$. 
\et

\begin{proof}
In direct analogy with the proof of Proposition \ref{relen} we have
\bea
\frac{1}{L}H\left(\p_{\L,N} ;\nu^{\L}_{\phi}\right) =-\frac1L \log\nu^{\L}_{\phi} [\Sigma_L =N]\to 0\ .
\eea
Convergence follows from the local limit theorem for triangular arrays (LLT) \cite{mcdonald1} since all marginals $\nu_\phi^x$ have exponential moments and $R(\phi )=\rho$.\\
For the relative entropy of a finite marginal on $\Delta$ we can write, using shorthands of the type $X_{\Delta ,k} =\{\Sigma_\Delta =k\}$,
\bea
\lefteqn{H\left(\p_{\L,N}^\Delta ;\nu^{\D}_{\phi}\right) =}\nonumber\\
& &\quad\sum_{k=0}^\infty \nu_\phi^\Delta (X_{\Delta ,k})\,\underbrace{\1_{k\leq N}\,\frac{\nu_\phi^{\Lambda\setminus\Delta} (X_{\Lambda\setminus\Delta ,N-k})}{\nu_\phi^\Lambda (X_{\Lambda ,N})}\log\frac{\nu_\phi^{\Lambda\setminus\Delta} (X_{\Lambda\setminus\Delta ,N-k})}{\nu_\phi^\Lambda (X_{\Lambda ,N})}}_{:= F_L (k)}\ ,
\eea
taking the form of an expectation of the function $F_L :\N\to\R$. We can again apply the LLT for every fixed $k$ and since $\Delta$ is finite with $N/(L-|\Delta |)\to\rho$ we have
\bea
\frac{\nu_\phi^{\Lambda\setminus\Delta} (X_{\Lambda\setminus\Delta ,N-k})}{\nu_\phi^\Lambda (X_{\Lambda ,N})}\to 1\ ,
\eea
which implies $F_L (k)\to 0$ pointwise as $L\to\infty$. Furthermore, $F_L (k)$ is bounded below by $-1/e$ and above by $CF_L (0)$ for $L$ large enough, again by the LLT. Therefore dominated convergence implies that $H\left(\p_{\L,N}^\Delta ;\nu^{\D}_{\phi}\right)\to 0$ as $L\to\infty$.\\
The above argument also immediately implies that the Radon-Nikodym derivative
\bea
h_\Lambda^\Delta (\eta^\Delta )=\frac{\nu_\phi^\Delta (\eta^\Delta )}{\pi_{\Lambda ,N}^\Delta (\eta^\Delta )}=\1_{\Sigma_\Delta \leq N} (\eta^\Delta )\,\frac{\nu_\phi^{\Lambda\setminus\Delta} (\Sigma_{\Lambda\setminus\Delta} =N-|\eta^\Delta |)}{\nu_\phi^\Lambda (\Sigma_{\Lambda} =N)}
\eea
is bounded. So analogously to the proof of Theorem \ref{equi} for homogeneous systems this implies convergence for $L^{1+\epsilon}$-integrable cylinder functions.\\
If $\nu_{\phi_c}$ has finite second moments they are uniformly bounded by (\ref{wbound}) and the same LLT applies for both parts of the proof.
\end{proof}

\subsection{Equivalence for supercritical systems}

If we know that the condensed phase is delocalized with $D_\phi =[0,\phi_c ]$ we have convergence of the full specific relative entropy in analogy to the homogeneous result in Proposition \ref{relen}. Otherwise, the critical measure $\nu_{\phi_c}$ does not exist on the full lattice, and we have to focus our attention to the fluid phase to show equivalence. We first give a result on the purely localized case under additional assumptions, and formulate a general, weaker result at the end of this section.

\bt{thm:Nonhomo-equiv}
Consider a uniformly bounded sequence $\lambda_1 ,\lambda_2 ,\ldots$
\revs{and $\rho_c <\infty$ as defined in (\ref{rhoc}). Then under assumption (\ref{werner}):\revf}
\begin{enumerate}
	\item \textbf{Delocalized case.} If $\lambda_x <1/\phi_c$ for all $x\in\N$, and the critical measure $\nu_{\phi_c}$ has finite second moments we have for all $\rho \geq\rho_c$
    \begin{align}
      \frac{1}{L}H\left(\p_{\L,N} ;\nu^{\L}_{\phi_c}\right)
      \to 0\,, \quad \textrm{as $L\to \infty$ and $N/L \to \rho$\ .}
    \end{align}
  \item \textbf{Localized case.} If $\Delta =\{ x:\lambda_x =1/\phi_c \}\neq\emptyset$ and for all $y\not\in\Delta$, $1/\lambda_y >\phi_c +\delta$ for some $\delta >0$, we have for all $\rho\geq \rho_c$
      \begin{align}\label{lcase1}
      \frac{1}{L}H\left(\p_{\L,N}^{\L\setminus\D} ;
        \nu^{\L\setminus \D}_{\phi_c}\right) \to 0, \quad \textrm{as $L\to  \infty$ and $N/L \to \rho$\ .}
      \end{align}
			Furthermore, the volume fraction of the condensed phase vanishes,\quad $|\Delta \cap\Lambda |/L\to 0$ as $L\to\infty$.\\
\end{enumerate}
\et

\begin{proof}
\textit{1. Delocalized case.}\\
Using again the representation \eqref{ldcon} of specific relative entropy, we have
  \begin{align*}
    \frac{1}{L} H\left(\p_{\L,N} ; \nu^{\L}_{\Phi(\rho)}\right) =
    -\frac{1}{L}\log \nu^{\L}_{\Phi(\rho)}\left[\Sigma_\Lambda =N\right]\,.
  \end{align*}
Define a sequence of `slowest' sites
  $$x_L :=\min \big\{ x\in\Lambda :\lambda_x =\max_{y\in\Lambda} \lambda_y \big\}\ ,$$
  i.e. smallest possible site indidces where the maximum of $\lambda$ is attained in $\Lambda$. With (\ref{domlim}) we have $\lambda_{x_L} \nearrow 1/\phi_c$ and since $\phi_c \in D_\phi$ the $\lambda_x$ do not attain their supremum and $x_L \to\infty$ as $L\to\infty$.
  
  As in the proof of Proposition \ref{relen} for $\rho > \rho_c$ we give an upper bound on the specific relative
  entropy by distributing the entire excess mass on the site $x_L$,
\bea\label{paff}
    -\frac{1}{L}\log \nu^{\L}_{\phi_c}\left[\Sigma_\Lambda =N\right] &\leq &
    -\frac{1}{L}\log \bar\nu^{x_L}_{\phi_c}(K_L ) \nonumber\\
    & &-\frac{1}{L}\log
    \nu^{\L\setminus \{x_L\}}_{\phi_c}\left[\Sigma_{\L\setminus
        \{ x_L\}} =N-K_L\right] 
\eea
  where $K_L := \lceil N - \rho_c L \rceil$. The first term is given by
  \begin{align*}
    -\frac{1}{L}\log \bar\nu^{x_L}_{\phi_c}(K_L ) =-
    \frac{K_L }{L}\log (\phi_c \l_{x_L} ) - \frac{1}{L}\log w_{x_L} (K_L ) + \frac{1}{L}\log
    z_{x_L} (\phi_c )
  \end{align*}
and each contribution vanishes as $L\to\infty$: For the first contribution $K_L /L\leq\rho$ and $\l_{x_L} \nearrow 1/\phi_c$, and the second term is bounded above by $-\frac{1}{L}\log w_- (K_L )$ which vanishes with (\ref{wbound}). The third term characterizes the limiting contribution of the condensed phase to the critical pressure $p(\phi_c )$ (\ref{pressure}) which can be written as
  \bea
  p(\phi_c )=\lim_{L\to\infty} \Big(\frac1L \sum_{x\in\Lambda\setminus\Delta} \log z_x (\phi_c )+\frac1L \sum_{x\in\Lambda\cap\Delta} \log z_x (\phi_c )\Big)\ ,
  \eea
  where $\Delta =\{ x_L :L=1,2\ldots\}$ is the trace of the sequence $x_L$. Note that $|\Lambda\cap\Delta |\to\infty$ and therefore $\frac{1}{L}\log
    z_{x_L} (\phi_c )\to 0$, since otherwise the second contribution to the pressure would diverge, contradicting $\phi_c \in D_\phi$.\\
  Since $(N-K_L )/L \to \r_c$ and since the critical measure $\nu_{\phi_c}$ has finite
  second moments which are then uniformly bounded by (\ref{wbound}) we may apply again the LLT (see Appendix) so that
  the second term of (\ref{paff}) vanishes in the limit, which completes the proof of case 1.\\

\textit{2. Localized case.}\\
$\nu_{\phi_c}$ does not exist on $\Lambda$ but we can use a different reference measure to write
$\pi_{\Lambda ,N} =\nu_{\phi_c -\epsilon} [\,.\, |\Sigma_\Lambda =N]$ as a conditional distribution for any $\epsilon\in (0,\phi_c)$. 
Then we get
  \begin{gather}
    \frac{1}{L}H\left(\p_{\L,N}^{\L\setminus\D} ;
        \nu^{\L\setminus \D}_{\phi_c}\right) = -\frac{1}{L}\log \nu^\L_{\phi_c - \e}[\Sigma_{\L} = N] + 
      \frac{1}{L}\log \frac{z^{\L\setminus\D}(\phi_c)}{z^{\L\setminus\D}(\phi_c-\e)} +    \nonumber \\
      \quad +\pi_{\L,N}\left( \frac{N {-} \Sigma_\D}{L} \right)\log\frac{\phi_c {-}\e}{\phi_c}
      {+} \frac{1}{L}\sum_{k=0}^N\pi_{\L,N} [\Sigma_\Delta {=}k]\log\nu^\D_{\phi_c - \e}[\Sigma_\D {=}k]\nonumber\\
      \quad\leq -\frac{1}{L}\log \nu^\L_{\phi_c - \e}[\Sigma_{\L}(\eta) = N] +\frac{1}{L}\log \frac{z^{\L\setminus\D}(\phi_c)}{z^{\L\setminus\D}(\phi_c-\e)} \,,    \label{eq:s1}
  \end{gather}
  where the upper bound follows immediately since the final two terms are negative.
  Since on $\L\setminus \D$ we have $1/\l_x > \phi_c + \d$, $\lim\limits_{L\to\infty}\frac{1}{L}\log
  z^{\L\setminus\D}(\phi)<\infty$ for all $\phi\in [0,\phi_c +\delta /2)$, 
  and as a limit of convex functions it is convex and therefore Lipschitz continuous (see \cite[Appendix A]{dembozeitouni}).
  So if we choose $\epsilon$ small enough we have
  \begin{align*}
 \lim_{L\to\infty}\frac{1}{L}\log \frac{z^{\L\setminus\D}(\phi_c)}{z^{\L\setminus\D}(\phi_c-\e)}\leq \wt\e
  \end{align*}
for any given $\wt\e >0$.\\
  For an upper bound on the first term we again consider only the event that the excess mass is all distributed on the `slowest' site $x_L$. Completely analogously to estimating (\ref{paff}) in case 1 we get
  \begin{align}
    \label{eq:s2}
    -\lim_{L\to\infty} \frac{1}{L}\log \nu^\L_{\phi_c - \e}[\Sigma_{\L}(\eta) = N] \leq -\rho\log\frac{\phi_c -\e}{\phi_c}\leq \wt\e\ ,
  \end{align}
  if we choose $\e$ small enough. Thus we can show that for all $\wt\e >0$
$$
\lim_{L\to\infty} \frac{1}{L}H\left(\p_{\L,N}^{\L\setminus\D} ;\nu^{\L\setminus \D}_{\phi_c}\right) <2\wt\e
$$
and therefore the specific relative entropy vanishes for all $\delta$.\\
An extensive volume fraction $|\Delta\cap\Lambda|$ would imply $\rho_c =\infty$, since it is defined in (\ref{rhoc}) by the left limit of $R$ (\ref{limdens}), and the contribution on $\Delta$ would diverge since $R_x (\phi_c ) =\infty$ for all $x\in\Delta$ by assumption (\ref{werner}).
\end{proof}
%
%
%
Without the additional assumption (\ref{werner}) one can formulate a slightly weaker result that applies in general, by excluding a set that is potentially larger than the condensed phase.

\bt[General case.]{general}
Consider a uniformly bounded sequence $\lambda_1 ,\lambda_2 ,\ldots$ and assume $\rho_c <\infty$ as defined in (\ref{rhoc}).
Define
  \bea\label{ddef}
  \Delta =\Delta_\delta :=\Big\{ x\in \N : 1/\l_x < \phi_c + \delta\ \mbox{and}\ R_x (\phi_c )>1/\delta\Big\}\ .
  \eea
  Then, for any $\delta >0$ and $\rho\geq \rho_c$ we have
      \begin{align}\label{lcase}
      \frac{1}{L}H\left(\p_{\L,N}^{\L\setminus\D} \mid
        \nu^{\L\setminus \D}_{\phi_c}\right) \to 0, \quad \textrm{as $L\to  \infty$ and $N/L \to \rho$\ ,}
      \end{align}
  and\quad $\displaystyle r(\delta ):=\lim_{L\to\infty} \frac{|\Lambda\setminus\Delta|}{L}\to 1$\quad as $\delta\to 0$.\\
\et

Note that for large $\delta$ we can have $\Delta =\Lambda$ and (\ref{lcase}) holds trivially. The interesting case is that it holds also for arbitrarily small $\delta$, where the volume fraction of the fluid phase approaches $1$. To ensure this in general we need the second condition on the density in (\ref{ddef}), since otherwise the set $\Delta$ could be extensive and the condensate be delocalized somewhere within that set.

\begin{proof}
Assume that $\delta >0$ is small enough so that $\Delta\neq\N$, otherwise there is nothing to show. Note that the critical measure $\nu_{\phi_c}^{\Lambda\setminus\Delta}$ outside $\Delta$ is well defined for all $\delta >0$, since either $R_x (\phi_c )\leq 1/\delta$ or the tails of the marginals have exponential moments. For each fixed $\delta$ we can proceed exactly analogous to the proof of Case 2 in Theorem \ref{thm:Nonhomo-equiv}, and conclude that the relative entropy density vanishes.\\
To estimate the volume fraction of the fluid phase $r(\delta ):=\lim_{L\to\infty} |\Lambda\setminus\Delta |/L$ note first that the limit exists for all $\delta >0$ due to subadditivity, and the right limit $r(0+)\in [0,1]$ exists since it is a monotone decreasing function. If $r(0+)<1$, then the set
\bea
\Delta_0 =\{ x:\lambda_x =1/\phi_c :R_x (\phi_c )=\infty\}
\eea
would cover a finite fraction of all sites, which would imply $\rho_c =\infty$.
\end{proof}

For supercritical systems, convergence of relative entropy of finite marginals cannot be concluded as easily from the local limit theorem as for subcritical systems. In addition, for inhomogeneous systems subadditivity of relative entropy also does not provide a simple bound as for homogeneous systems in Corollary \ref{gencon}. It remains an interesting open problem at this stage whether our results for the specific relative entropy can be used in general to imply convergence of marginals.

\section{Discussion\label{sed:discussion}}

We conclude the paper with a short discussion of further rigorous results on condensation in \revs{closed\revf} stochastic particle systems, which mostly focus on zero-range processes so far. This is clearly the richest model class, which can exhibit condensation due to particle interactions or spatial effects, or in other scaling limits as discussed below. The recently introduced explosive condensation model \cite{waclawetal11}, also discussed in more detail in Section \ref{sec:genres}, has the same rich structure in terms of stationary product measures and poses interesting questions for future work with respect to the dynamics. For the other two models mentioned in Section \ref{sec:defdyn}, the target process \cite{target} has only a restricted set of stationary product measures and further progress is very challenging, and the inclusion process exhibits homogeneous condensation only in a particular scaling limit with system size dependent parameters \cite{getal11}, which is still under investigation \cite{redigetal12}.

\subsection{Further stationary results\label{sec:further}}

Systems that exhibit condensation in the thermodynamic limit usually show the same phenomenon already on finite lattices, in the limit of a diverging number of particles or density. This has been studied in \cite{ferrarietal07} for homogeneous systems of i.i.d. random variables with regularly varying tails, including power laws that arise for zero-range dynamics of the form (\ref{zrconex}) with $\gamma =1$. It is shown that under canonical distributions $\pi_{\Lambda ,N}$ in the limit $N\to\infty$ the occupation number $M_\Lambda$ of the maximally occupied site diverges, whereas the rest of the system converges to a product measure with cricital marginals $\nu_{\phi_c}^1$ and density $\rho_c$. By symmetry, the location of the condensate (maximum) is chosen uniformly at random. The analogous result for inhomogeneous systems has recently been formulated in \cite{getal11}, where the condensate is located at the maximum of the the profile of the harmonic functions $\lambda_x$ (cf. Theorem \ref{spm}). In both cases, convergence to the product measure holds in distribution, and there is a strong law of large numbers for the condensate, i.e. $M_\Lambda /N\to 1\ a.s.$ as $N\to\infty$.

Another interesting scaling law concerns a detailed analysis of the thermodynamic limit at the critical density to study the onset of condensation and the nature of the transition. In \cite{agl} product measures arising for zero-range dynamics of the form (\ref{zrconex}) have been studied in the limit $N,L\to\infty$, $N/L\to\rho_c$ with an excess mass of subextensive order $o(L)$. It turns out that the condensate forms suddenly on a critical scale $N-\rho_c L\sim D_L$, which depends on the tail of the critical marginal $\nu_{\phi_c}^1$ characterized by the parameter $\gamma$ (\ref{zrconex}). It is of order $D_L \sim L\log L$ in the power law case with $\gamma =1$, and of order $L^{1/(1+\gamma)}$ in the stretched exponential case $\gamma\in (0,1)$. A law of large numbers for the excess mass fraction in the maximum is established, which jumps at the critical scale from zero to one in the power law, and to a positive value smaller than one in the stretched exponential case, where the excess mass is shared between bulk and condensate. Distributional limits for the fluctuations of the maximum are also covered, which change from standard extreme value statistics to Gaussian when the density crosses the critical scale. Results on fluctuations in the bulk show that the mass outside the maximum is distributed homogeneously. Some of these aspects of the crossover from sub- to supercritical behaviour have previously also been studied in \cite{evansetal05b}.

The same phenomenon for the zero-range process (\ref{zrconex}) at the critical scale has been studied \revs{by the authors\revf} in \cite{chlebounetal10,paulthesis}, where the discontinuity is established as the leading order finite size effects in a rigorous scaling limit. Simulation results reveal a switching between metastable fluid and condensed states close to the critical point for $\gamma\in (0,1)$, which are characterized using a current matching argument and an extension of homogeneous states to supercritical densities. The latter lead to strong finite size effects where the canonical current overshoots its thermodynamic limit, and coexistence of condensed and homogeneous states at supercritical densities can be relevant in real systems of moderate size such as vehicular traffic. This phenomenon of current overshoot has been studied also before in \cite{angeletal04} in a zero-range process with a single defect site.

It is well known that system-size dependent interaction potentials that lead to long range interactions can be used to stabilize finite-size effects and metastability in the thermodynamic limit. In this spirit, a simple zero-range process with size-dependent jump rates has been studied in \cite{stefan2} which exhibits a discontinuous condensation transition. There exist metastable homogeneous states at all densities, and in a condensed state, the condensate contributes a macroscopic amount to the canonical entropy, which results in a non-equivalence of the canonical and the grand-canonical measures. The saddle point structure of the free energy landscape of this model is analyzed by rigorous large deviation techniques in \cite{paulthesis,paulinprep}, and reveals a further dynamic transition above the critical density, where the stationary dynamics of the condensed phase is expected to change. No dynamic results have been proven so far for this model, and we comment on work in progress below.

Condensation in systems with more than one particle species or conserved quantity has been studied in \cite{evanshanney03,evanshanney04,godreche07,grosskinsky08} in the context of two-species zero-range processes, for which stationary product measures only exist under specific assumptions on the jump rates. While condensates still concentrate on single lattice sites, these models exhibit a richer phase diagram with further structure within the condensed regime. Of particular interest are states where both species condense, which can occur independently or be the result of cross-correlations with resulting correlations also in the locations of the condensates.

\subsection{The dynamics of condensation\label{sec:dynamics}}

While the understanding of stationary properties of condensation in stochastic particle systems is fairly complete by now, much less is known about the dynamics of condensation on a rigorous level. In heterogeneous zero-range processes with constant jump rates $u_x (n)\equiv u_x$ for $n\geq 0$, one can use attractivity of the process and coupling techniques. This has been done in \cite{landim96} to obtain rigorous results in a hydrodynamic limit for zero-range processes with defect sites, where the condensed part of the configuration is described by dirac measures appearing on sites with low rates. In \cite{andjeletal00} coupling techniques are used to characterize convergence to the critical stationary measure for zero-range processes with random rates $u_x$, initialzed at supercritical densities. This is extended to zero-range processes with general non-decreasing random rates $u_x (n)$ in \cite{ferrarisisko}. Further related results on hydrodynamic limits in exclusion models with particle disorder can be found in \cite{benjaminietal96}.

For supercritical homogeneous processes, the location of the condensate $X(\eta )=\mathrm{argmax}\{\eta_x :x\in\Lambda\}$ is distributed uniformly on $\Lambda$ under the canonical measures $\pi_{\Lambda ,N}$. Therefore, one expects the condensate to move on slow time scales in the limit of large system sizes for particle systems with ergodic dynamics, such as zero-range processes. If the system exhibits a proper separation of time scales, different condensate locations can be identified with metastable states, and equilibration in each state is fast compared to the time scale of motion and leads to a Markovian limit process on the lattice.  
The first rigorous results on the stationary condensate dynamics in reversible zero-range processes have been obtained in \cite{beltranetal10,beltranetal12}. On a fixed, general lattice $\Lambda$ with single particle rates $p(x,y)$, it is shown that the maximum location $X(\eta )$ in a process with rates of the form (\ref{zrconex}), $\gamma =1$, converges on the time scale $N^{1+b}$ to a random walk $(Y_t :t\geq 0)$ concentrating on the sites of $\Lambda$ with the maximal value of the harmonic function $\lambda_x$. Precisely,
\bea
\Big( X\big(\eta_{N^{1+b} t}\big) :t\geq 0\Big) \to (Y_t :t\geq 0)\quad\mbox{as }N\to\infty
\eea
weakly in the Skorohod topology on path space, where the rates of the limit process are proportional to the capacities of a single particle with dynamics given by $p(x,y)$. The proof is based on a potential theoretic approach to metastability using precise estimates on capacities for reversible systems (see \cite{bovier} and references therein). This approach has recently been extended to non-reversible dynamics \cite{beltranetal12b} and applied to the totally asymmetric zero-range process \cite{landim12}.

An important aspect in these results is to show that the system equilibrates fast enough in a metastable state on the time scale of the effective dynamics. A simple renewal-type approach could be used on finite lattices in the above results, since the process visits every configuration associated with a metastable state increasingly often before switching to another state. This has recently been extended to weaker conditions on the mixing and relaxation times of the dynamics within a metastable state defined with reflecting boundary conditions \cite{beltranetal13}. This approach is particularly suitable to extend the above results to the thermodynamic limit $L,N\to\infty$ with $N/L\to\rho >\rho_c$. This is current work in progress \cite{armendarizetal12} for reversible zero-range processes with rates (\ref{zrconex}) on a one-dimensional geometry with periodic boundary conditions, where the limit dynamics is expected to be a L\'evy-type process on the unit torus. First results in the same direction have been obtained in \cite{neukirch} where capacity estimates for the zero-range process are given in the limit $L,N\to\infty$ with diverging densities $N/L\to\infty$.

Another interesting aspect with recent first rigorous results is the approach to stationarity from homogeneous initial conditions and the formation of the condensed phase. Heuristic results on the separation of time scales in zero-range processes \cite{stefan,godreche03,godrecheetal05} predict a coarsening behaviour, where clusters form locally and exchange particles through the bulk with large clusters gaining on the expense of smaller ones. First rigorous results in this direction for reversible zero-range processes \cite{jara} address this question on a fixed lattice $\Lambda$ with diverging particle number $N\to\infty$. In the same scaling limit, the coarsening dynamics and the stationary motion of the condensate have recently been established in \cite{redigetal12} for symmetric inclusion processes with a vanishing diffusion parameter $d=d_N \to 0$ as $N\to\infty$ (cf. Section \ref{sec:defdyn}). This scaling leads to an explicit two-scale structure of the process, and established convergence results could be applied to identify the generator of the limiting process. In contrast to zero-range dynamics, clusters are mobile on the coarsening time scale and exhibit an interesting particle exchange dynamics via common empty nearest-neighbour sites, which can also lead to a spontaneous merge of two clusters. An extension to asymmetric dynamics seems feasible, and first heuristic results in related models \cite{waclawetal11} show a similar slinky motion and interaction of clusters, which are also observed in \cite{hirschberg} for non-Markovian zero-range dynamics. 
There are also heuristic results on hydrodynamic limits for condensed zero-range processes \cite{rosy}, which confirm the validity of the fundamental diagram in Figure \ref{fig2} in the full density range including $\rho >\rho_c$. This is clearly different from explosive condensation models \cite{waclawetal11} presented in the same figure, where the stationary current diverges with the system size for supercritical systems.


\subsection{Conclusion}

The aim of this paper was to provide an overview of rigorous results on condensation in stochastic particle systems, to illustrate these results with examples and embed them in the classical framework of phase transitions and the equivalence of ensembles for homogeneous and inhomogeneous systems. The presentation includes new results on relative entropy convergence and corollaries for convergence of test functions, as well as equivalence results for general inhomogeneous systems. While there are still some obvious open questions related to systems with inhomogeneities, in our view the most interesting fields of further study lie in the area of the dynamics of condensation. Particularly interesting questions include a hydrodynamic limit for supercritical processes including the dynamics of the condensed phase, a rigorous description of the coarsening dynamics in the hydrodynamic limit for symmetric and asymmetric processes, or whether there are attractive, homogeneous particle systems that exhibit condensation and allow an analysis with coupling techniques.

\appendix 
\section*{Appendix: Local limit theorems}
\setcounter{section}{1}
In this appendix we state relevant limit theorems for triangular arrays of independent non identical random variables that are key to results on the equivalence of ensembles for spatially inhomogeneous systems.

Details and a proof of the Lindeberg-Feller central limit theorem can be found in, for example, \cite{DurrettProbability}.
The local central limit theorem can be found in \cite{mcdonald1} and \cite{mcdonald2}.
For each $L$, let $\xi_{x,L}$, $1\leq x \leq L$, be independent non identical random variables whose law depends on the $x$ and the number of random variables $L$ (a triangular array of random variables).

\begin{thm}[The Lindeberg-Feller central limit theorem]\mbox{}\\
Suppose $\mathbb{E}[\xi_{x,L}] = 0$, and
\begin{enumerate}
\item[(i)] $\sum\limits_{x=1}^L \mathbb{E}[\xi_{x,L}^2] \to 1$ as $L\to\infty$.
\item[(ii)] For all $\epsilon>0$, $\sum\limits_{x=1}^{L}\mathbb{E}\left[\left\vert\xi_{x,L}\right\vert^2\ind_{\vert \xi_{x,L}\vert > \epsilon}\right] \to 0$ as $n\to\infty$.
\end{enumerate}
Then $\sum\limits_{x=1}^L\xi_{x,L}$ converges in distribution to the standard normal as $n\to\infty$.
\end{thm}

For example we typically apply the central limit theorem to the centered and standardized sum of the single site occupations under the grand canonical measures, i.e. $$\xi_{x,L} = \frac{\h_x-R_x(\phi)}{\sqrt{L\,\sum_y\textrm{Var}_\phi(\h_y)}}$$ where $\h_x$ has law $\nu^x_{\phi}$ not depending on the system size $L$.
Then $(i)$ follows by definition and $(ii)$ follows by a dominated convergence argument if the second moments are uniformly bounded.

Now we assume that $\h_{x,L}$, $1\leq x \leq L$, is a triangular array of independent integer valued random variables where $\h_{x,L}$ has law  $P_{x,L}$.
The Bernoulli part decomposition of the random variables $\h_{x,L}$ is expressed in terms of,
\[
q(P_{x,L}) = \sum_{n}\left( P_{x,L}[n] \wedge P_{x,L}[n+1] \right)\ .
\]
Define\quad $Q_L = \sum_{x=1}^L q(P_{x,L})$\ .

\begin{thm}[Local central limit theorem \cite{mcdonald1}]\label{thm:LLT}\mbox{}\\
Let $\Sigma_L = \sum_{x=1}^L\h_{x,L}$. Suppose there exist sequences $B_L > 0$ and $A_L$, $L\geq 1$, such that $B_L\to\infty$, $\limsup B_L^2/Q_L < \infty$, and $(\Sigma_L-A_L)/B_L$ converges in distribution to the standard normal. Then,
\begin{align}
\sup_{n}\left\vert B_L P[\Sigma_L = n] - \Phi\left(\frac{n-A_L}{B_L}\right)\right\vert\to 0 \quad \textrm{as} \quad L\to\infty\ ,
\end{align}
where $\Phi$ is the standard normal density.
\end{thm}

An alternative form of the local limit theorem can be found in \cite{mcdonald2}. in our cases $A_L \sim L$ and $B_L \sim \sqrt{L}$, and the main condition is to show that $Q_L \sim L$. In fact, using the structure of the marginals $(\phi\lambda_x )^n w_x(n)$ with uniform regularity of the $w_x (n)$ (\ref{wbound}) it is easy to see that in fact $Q_L \sim L$.

\section*{Acknowledgements}
S.G. acknowledges support by the Engineering and Physical Sciences Research
Council (EPSRC), Grant No. EP/I014799/1. \revs{P.C. acknowledges support and funding from the University of Warwick as an IAS Global Research Fellow\revf}.
We are grateful for inspiring discussions with colleagues, in particular E. Saada, T. Gobron, M.R. Evans, F. Redig and C. Godr{\`e}che.

%

\end{document}